\documentclass[aps,preprint,superscriptaddress,onecolumn,12pt,nofootinbib]{revtex4}
\usepackage{amsfonts}
\usepackage{bm}
\usepackage{soul}
\usepackage{makecell,multirow}
\usepackage{rotating}
\usepackage[utf8]{inputenc}
\usepackage{amsmath}
\usepackage{hyperref}
\hypersetup{
    colorlinks=true,
    urlcolor= blue,
    citecolor=blue,
    linkcolor= blue}
\usepackage{makecell}

\usepackage{amssymb}
\usepackage{tensor}
\usepackage{graphicx}%
\usepackage{bigints}
\usepackage{bbm}
\usepackage{graphicx}
\usepackage{subfigure}
\usepackage{epsf}
\usepackage{bm}
\usepackage{amsmath}
\usepackage{amsfonts}
\usepackage{amssymb}
\usepackage{graphicx}
\usepackage{tabularx}
\usepackage{multirow}
\usepackage{color}%
\providecommand{\U}[1]{\protect\rule{.1in}{.1in}}
\def\d{\mbox{\rm d}}
\def\e{\mbox{\rm e}}

\newcommand{\ie}{\begin{equation}}
\newcommand{\fe}{\end{equation}}

\newcommand{\mincir}{\raise
-3.truept\hbox{\rlap{\hbox{$\sim$}}\raise4.truept\hbox{$<$}\ }}
\newcommand{\magcir}{\raise
-3.truept\hbox{\rlap{\hbox{$\sim$}}\raise4.truept\hbox{$>$}\ }}

\providecommand{\U}[1]{\protect\rule{.1in}{.1in}}

\usepackage{tikz,xcolor,hyperref}

\definecolor{lime}{HTML}{A6CE39}
\DeclareRobustCommand{\orcidicon}{%
	\begin{tikzpicture}
	\draw[lime, fill=lime] (0,0) 
	circle [radius=0.16] 
	node[white] {{\fontfamily{qag}\selectfont \tiny ID}};
	\draw[white, fill=white] (-0.0625,0.095) 
	circle [radius=0.007];
	\end{tikzpicture}
	\hspace{-2mm}
}

\foreach \x in {A, ..., Z}{%
	\expandafter\xdef\csname orcid\x\endcsname{\noexpand\href{https://orcid.org/\csname orcidauthor\x\endcsname}{\noexpand\orcidicon}}
}

\begin{document}

\title{\Large{On the information behavior from quadratically coupled accelerated detectors}}

\author{P. H. M. Barros\orcidA{}}
\email{phmbarros@ufpi.edu.br (The corresponding author)}

\affiliation{Departamento de F\'{i}sica, Universidade Federal do Piau\'{i} (UFPI), Campus Min. Petr\^{o}nio Portella - Ininga, Teresina - PI, 64049-550 - Brazil}

\author{P. R. S. Carvalho\orcidB{}}
\email{prscarvalho@ufpi.edu.br}
\affiliation{Departamento de F\'{i}sica, Universidade Federal do Piau\'{i} (UFPI), Campus Min. Petr\^{o}nio Portella - Ininga, Teresina - PI, 64049-550 - Brazil}


\author{H. A. S. Costa\orcidC{}}
\email{hascosta@ufpi.edu.br}
\affiliation{Departamento de F\'{i}sica, Universidade Federal do Piau\'{i} (UFPI), Campus Min. Petr\^{o}nio Portella - Ininga, Teresina - PI, 64049-550 - Brazil}


\begin{abstract}

In this work, we propose to investigate the information behavior of quantum systems through accelerated detectors quadratically coupled with a massless scalar field. In addition, we made detailed comparisons with the case of linear coupling. The perturbative method was used to evolve the density matrix that describes the interaction of the detector-field system during a finite time. The systems studied were: accelerated single-qubit, quantum interferometric circuit, and the which-path distinguishability circuit. The results on the probability transition rates show that quadratic coupling amplifies the Unruh effect. This is due to the modification of the interaction structure, allowing the simultaneous absorption of multiple quanta. Our findings showed that the information is degraded more quickly in the case of quadratic coupling, when compared to the linear case. Furthermore, this change is mainly given by the coupling constant and by an additional factor that arises in the case of quadratic coupling. Therefore, these results indicate that the nature of the coupling between the detector and the field plays a fundamental role in the behavior of quantum information in high acceleration regimes.


\end{abstract}

\maketitle


	
\section{Introduction}

Modern theoretical physics continues to uncover deeply counterintuitive phenomena at the intersection between quantum mechanics and general relativity. One of the most intriguing examples is the so--called Fulling--Davies--Unruh effect \cite{fulling1973, Davies_1975, unruh1976}, according to which a uniformly accelerated observer in quantum vacuum experiences thermal radiation as if immersed in a particle bath -- an absent phenomenon for inertial observers \cite{Crispino2008}. This prediction, formulated by William Unruh in 1976 \cite{unruh1976}, demonstrates that the particle detection is connected to the presence of a Rindler horizon in the accelerated reference frame, which is mathematically equivalent to the horizon of black holes \cite{birrell1984quantum}.
This implies that different quantization procedures produce ambiguous definitions of particle numbers \cite{grove1983notes, wald1994quantum, lancaster2014quantum}.

Aiming to formally investigate this effect and other quantum aspects of field theory in inertial reference frames, the Unruh--DeWitt (UDW) detector was proposed \cite{unruh1976, DeWitt1980}: a simplified model that represents a two-level system coupled to a quantum field. Although simple, this model captures the essence of how different observers interact with quantum fields. The UDW detector serves as a conceptual instrument for probing and quantifying the particles generated as a consequence of acceleration. In this way, it can be said that particles are defined by what a particle detector detects \cite{davies1984}.

In recent years, relativistic quantum information (RQI) has become a new research area that combines gravitational physics and quantum computing. It seeks to explore the role of relativistic effects in quantum information processing protocols \cite{martin2011relativistic, Mann2012}. Even when excluding approaches to quantum gravity, RQI encompasses a broad range of investigations that differ significantly in scope, namely: the use of quantum probes to study the Unruh and Hawking effects \cite{hawking1975particle, unruh1976, DeWitt1980, wald1994quantum}, as well as the development of relativistic quantum communication and computation protocols \cite{Adlam2015, Landulfo2016, Martinez2015, Martinez2016, Martinez2020, Martinez2021, Tjoa2022, Lapponi2023}. Besides, Unruh's work \cite{Unruh1981Experimental} brought attention to the thermal emission that is expected from black holes. Showing that analogous systems, such as sonic horizons in fluids, also predict similar thermal emission.

A variety of methodologies have been proposed for the detection of particles associated with the Unruh effect. Among these, the quantum coherence approach has gained increasing attention \cite{tian2012unruh, Wang2016, he2018multipartite, Nesterov2020, Zhang2022, huang2022, Harikrishnan2022, xu2023decoherence, Pedro2024robustness}. In this framework, a UDW detector interacts with the quantum field by absorbing particles, which alters its eigenstates and results in a loss of quantum coherence. Recent studies have identified specific conditions that amplify this coherence degradation. For example, such effects have been observed when the vacuum is treated as a dispersive medium \cite{barros2024dispersive}, and similarly in the presence of a gravitational wave background \cite{barros2024detecting}. In contrast, our recent results indicate that the mass of the scalar field may provide a protective mechanism \cite{pedro2025mitigating}, mitigating the information degradation due to the Unruh effect.

Another approach for detecting this effect involves the use of quantum interferometric circuits \cite{Costa2020interferometry, GoodingUnruh2020interferometric, Lopes2021interference, Pedro2024robustness, barros2024detecting, pedro2025mitigating}, where the created particles exhibit modified probability amplitudes, reduced visibility, and altered interference patterns.
Moreover, the interferometric setup can be adapted to probe particle-like behavior by analyzing which-path distinguishability \cite{Jaeger1995, Englert1996, miniatura2007path, bera2015duality}. As a result, it becomes possible to derive the complementarity relation by jointly considering wave-like and particle-like characteristics. This framework paves the groundwork for investigating how the Unruh effect influences wave-particle duality \cite{Pedro2024robustness, pedro2025mitigating}.

It is worth noting that the vast majority of studies involving the UDW detector were done using a linear coupling of the detector with the quantum field. Besides, there are some studies that considered the quadratic coupling \cite{hinton1984particle, takagi1985response, Takagi1986vacuum}, they realized that the detector responds differently from the case coupled linearly.
Additionally, the quadratic coupling was considered in various scenarios, namely: fermonic and bosonic fields linearly and quadratically coupled \cite{MartinezRenormalized2016} (presenting a renormalization method), entanglement harvesting and divergences in quadratic UDW detector pairs \cite{Sachs2017entanglement}, scalar and fermionic Unruh Otto engines \cite{gray2018scalar}, entanglement of two accelerated detectors with an approach different \cite{wu2023birth}, response of accelerated and nonlinearly coupled detectors \cite{sriramkumar2001response}, enhancement of an UDW battery performance through quadratic coupling \cite{mukherjee2024enhancement}, and others.

Motivated by this, in this work we will study the influence of quadratic coupling on the information of accelerated quantum systems, and compare with the case of linear coupling. We structure our work as follows: In Sect. \ref{sec:2}, we studied UDW's theory, for a quadratically coupled detector with a massless scalar field, we find the Wightman function and transition probability rates. Next, in Sect. \ref{sec:3}, we investigate how quadratic coupling affects the information of some quantum systems, such as: accelerated single qubit, quantum interferometric circuit and which-path distinguishability circuit. Finally, in Sect. \ref{sec:4}, we present a summary of our findings as well as our conclusions compared to the case of linear coupling.

For convenience, we adopt natural units where $c = \hbar = k_B = 1$ and the metric signature is $(+,-,-,-)$ throughout the body of this paper.


\section{Unruh--DeWitt theory with quadratic coupling\label{sec:2}}

\subsection{The theoretical framework\label{Sec IIA}}

For the purposes of our study, we begin by considering an UDW detector, that is, a point-like detector characterized by two energy levels: a ground state $\vert g\rangle$ and an excited state $\vert e\rangle$. In this framework, it is considered that $\tau$ is the proper time of the detector traveling through a world line given by $x(\tau)$. Let us consider the scenario where the detector interacts quadratically with a massless scalar field given by $\phi[x(\tau)]$ via a detector monopole moment $\mu(\tau)$. Thus, the Hamiltonian that represents the quadratic interaction between the detector and the scalar field is expressed as follows:
\begin{eqnarray}
  \mathcal{H}_{\mathrm{int}} = \lambda_{\phi^2}\chi(\tau)\mu(\tau) \otimes \phi^2[x(\tau)],
  \label{Hint}
\end{eqnarray}
where $\lambda_{\phi^2}$ is the strength of the interaction, and $\chi(\tau)$ represents the switching function, which governs the activation and deactivation of the detector. Given the finite duration of interaction, this function satisfies the following conditions: $\chi(\tau) \approx 1$ for $|\tau| \ll T$, and $\chi(\tau) \approx 0$ for $|\tau| \gg T$, where $T$ characterizes the interaction time.

Before proceeding with the construction of the theoretical framework of this work, it is important to emphasize one last point. As observed by \cite{gray2018scalar}, for each coupling used we will have a different dimensionality, on the other hand, the same occurs with each coupling constant. In fact, for dimensionless switching functions, it can be shown that $[\lambda_{\phi^2}] \equiv -\Delta_{\phi^2} = 3 - d$ (where $d$ is the dimension) considering the quadratic coupling with scalar fields, in units where $[x] = 1$. However, it is possible to write a dimensionless constant of the type $\Lambda = \lambda_{\phi^2}/\xi^{\Delta_{\phi^2}}$, where $\xi$ is a time scale of any perturbative approximation of the detector evolution (for more details see \cite{gray2018scalar}).
For the case of this work, we have that $d=4$ and the detector evolution time scale is $\xi \sim
 \Omega^{-1}$, and therefore we obtain, $\Lambda = \lambda_{\phi^2}\Omega$.

The interaction should promote transitions in the detector from the energy state $g$ to the energy state $e$, while in the field, the interaction promotes a transition from the vacuum $\vert 0_\mathcal{M} \rangle$ to a particle state $\vert \Psi \rangle$. Furthermore, by perturbation theory up to first order in $\lambda_{\phi^2}$, the probability amplitude of the interaction occurring is given by
\begin{eqnarray}
    \mathcal{A}_{\phi^2} = i\lambda_{\phi^2} \langle e,\Psi\vert \int^{+\infty}_{-\infty} \mu(\tau)\chi(\tau) \phi^2[x(\tau)] \d\tau \vert 0_{\mathcal{M}},g \rangle.
    \label{prob_amplitude}
\end{eqnarray}
It is important to emphasize that the monopole momentum operator $\mu(\tau)$ has the function of connecting the ground state and the excited state of the detector. Furthermore, it depends only on the detector, not on the scalar field in question. Besides, using the Heisenberg representation, we have that the detector's monopole moment operator evolves in time as
\begin{eqnarray}
    \mu(\tau) = \e^{i\mathcal{H}_0\tau} \mu(0)\e^{-i\mathcal{H}_0\tau},
    \label{monopole_moment}
\end{eqnarray}
where $\mathcal{H}_0$ is the Hamiltonian of the detector, i.e., $\mathcal{H}_0 \vert g\rangle = g \vert g\rangle$ and $\mathcal{H}_0 \vert e\rangle = e\vert e\rangle$. In this way, substituting Eq. (\ref{monopole_moment}) in Eq. (\ref{prob_amplitude}), we obtain
\begin{eqnarray}
    \mathcal{A}_{\phi^2}(\Omega) = i\lambda_{\phi^2} \langle e\vert \mu(0) \vert g\rangle \int^{+\infty}_{-\infty}\d\tau \chi(\tau)\e^{i\Omega\tau} \langle \Psi\vert \phi^2[x(\tau)] \vert 0_{\mathcal{M}}\rangle,
    \label{prob_amplitude2}
\end{eqnarray}
with $\Omega = e - g$ denoting the detector transition frequency. Thus, the probability of measuring the transition, or simply transition probability $\mathcal{P}(\Omega)$, is obtained by taking the square of Eq. (\ref{prob_amplitude2}) and summing over all possible final states $\vert \Psi \rangle$, in this way we obtain,
\begin{eqnarray}
    \mathcal{P}_{\phi^2}(\Omega) &=& \lambda_{\phi^2}^2 \vert\langle e\vert \mu(0) \vert g\rangle\vert^2 \nonumber\\
    &\times& \int^{+\infty}_{-\infty}\d\tau \int^{+\infty}_{-\infty}\d\tau' \chi(\tau)\chi(\tau')\e^{-i\Omega(\tau-\tau')} \langle 0_{\mathcal{M}}\vert :\phi^2[x(\tau)]::\phi^2[x(\tau')]: \vert 0_{\mathcal{M}}\rangle,
    \label{prob_transition}
\end{eqnarray}
where the sum over all possible final states allowed us to use the completeness relation $\mathbb{I} = \sum_{\Psi} \vert \Psi\rangle \langle \Psi\vert$. Note that in Eq. (\ref{prob_transition}) we use the standard field theoretic technique of normal ordering \cite{MartinezRenormalized2016}, where $:\hat{A}: = \hat{A} - \langle 0_{\mathcal{M}}\vert \hat{A} \vert 0_{\mathcal{M}} \rangle $ for an operator $\hat{A}$. The normal ordering technique is necessary because, when considering quadratic scalar and fermionic detector couplings there are persistent divergences that cannot be regulated either by the switching function or by smearing the detector \cite{MartinezRenormalized2016, Takagi1986vacuum, Louko2016}.

Therefore, aiming at a more simplified interpretation we can write the transition probability as follows,
\begin{eqnarray}
    \mathcal{P}_{\phi^2}(\Omega) = \lambda_{\phi^2}^2 \vert\langle e\vert \mu(0) \vert g\rangle\vert^2 \mathcal{F}_{\phi^2}(\Omega),
    \label{prob_transition2}
\end{eqnarray}
where $\mathcal{F}_{\phi^2}(\Omega)$ is known as the detector response function\footnote{This function carries with it the information of how the detector responds to the interaction with the quantum field. Additionally, it is noted that the first part of the right-hand side of Eq. (\ref{prob_transition}) depends only on the internal characteristics of the detector and the coupling with the scalar field.}, and is responsible for identifying the particles with the detector energy variation $\Omega$ created due to the interaction with the quantum field. The response function is written as 
\begin{eqnarray}
    \mathcal{F}_{\phi^2}(\Omega) = \int^{+\infty}_{-\infty}\d\tau \int^{+\infty}_{-\infty}\d\tau' \chi(\tau)\chi(\tau')\e^{-i\Omega(\tau-\tau')} \mathcal{W}_{\phi^2}(x,x'),
    \label{Response_Function_Infty}
\end{eqnarray}
where
\begin{eqnarray}
    \mathcal{W}_{\phi^2}(x,x') = \langle 0_{\mathcal{M}}\vert :\phi^2[x(\tau)]::\phi^2[x(\tau')]: \vert 0_{\mathcal{M}}\rangle,
    \label{defWightman}
\end{eqnarray}
is called the Wightman function \cite{Wightman1956}, which corresponds to a correlation function. If the worldline along which the detector travels corresponds to integral curves of timelike Killing fields, then the Eq. (\ref{defWightman}) becomes invariant under time translations in the detector's frame of reference and depends only on the time difference \cite{letaw1981quantized, padmanabhan1982general}, i.e., $\mathcal{W}_{\phi^2}[x(\tau),x'(\tau)] = \mathcal{W}_{\phi^2}(\tau,\tau') = \mathcal{W}_{\phi^2}(\Delta\tau)$.

In this way, the double integral in Eq. (\ref{Response_Function_Infty}) reduces to a Fourier transform of $\mathcal{W}_{\phi^2}(\Delta\tau)$ multiplied by an infinite time interval that naturally causes $\mathcal{F}_{\phi^2}(\Omega)$ to blow up. In this case, we get around the divergence by working with the transition probability rate, given by
\begin{eqnarray}
    \mathcal{R}_{\phi^2}(\Omega) = \int^{+\infty}_{-\infty}\d(\Delta\tau)  \chi(\tau)\chi(\tau') \e^{-i\Omega\Delta\tau} \mathcal{W}_{\phi^2}(\Delta\tau),
    \label{Response_Rate}
\end{eqnarray}
instead of dealing with the probability itself, in other words, we deal with the response function per unit time. Thus, it is clear that in order to know how the detector responds to the quadratic interaction with the scalar field, we must first obtain an expression for the correlation function of the system in question.

\subsection{Wightman function}

This section consists of obtaining the Wightman function for a detector that has quadratic coupling and uniform acceleration. However, it is well known in the quantum information community in relativistic regimes the expression for the linear case considering the Kubo-Martin-Schwinger (KMS) condition \cite{kubo1957statistical, Martin1959Schwinger}. Aware of this, through the Eq. (\ref{defWightman2}) we have
\begin{eqnarray}
    \mathcal{W}_{\phi^2}(\Delta t,\Delta z) = 2[\mathcal{W}(\Delta t,\Delta z)]^{2}.
    \label{defWightman3}
\end{eqnarray}
However, applying these results along a trajectory suffering a constant and uniform acceleration $a$ in the direction $z$, for this, the intrinsic coordinate system within the detector frame is connected with the Minkowsky coordinates through the Rindler transformations \cite{rindler1966kruskal}, 
\begin{eqnarray}
    t = \frac{1}{a} \sinh{a\tau}, \quad z = \frac{1}{a} \cosh{a\tau}, \quad x = y = 0.
    \label{trajectory}
\end{eqnarray}
On the other hand, it is well known that for these coordinates the Wightman function for linear coupling is given by
\begin{eqnarray}
    \mathcal{W}(\Delta t,\Delta z) = -\frac{1}{4\pi^{2}}\frac{1}{(\Delta t^2 - \Delta z^{2} - i\epsilon)},
\end{eqnarray}
and using Eq.(\ref{defWightman3}) is obtained
\begin{eqnarray}
    \mathcal{W}_{\phi^2}(\Delta t,\Delta z) = \frac{1}{8\pi^{4}}\frac{1}{(\Delta t^2 - \Delta z^{2} - i\epsilon)^2},
    \label{func_wightman_quadratic}
\end{eqnarray}
and using the elementary properties of hyperbolic functions in Eq. (\ref{trajectory}), you get
\begin{eqnarray}
    \Delta t &=& \frac{2}{a} \cosh{\left( \frac{\tau+\tau'}{2a^{-1}}\right)} \sinh{\left( \frac{\tau-\tau'}{2a^{-1}}\right)}, \\
    \Delta z &=& \frac{2}{a} \sinh{\left( \frac{\tau+\tau'}{2a^{-1}}\right)} \sinh{\left( \frac{\tau-\tau'}{2a^{-1}}\right)}.
\end{eqnarray}
Replacing these expressions in the Eq. (\ref{func_wightman_quadratic}) and using the hyperbolic property $\cosh^2{x} - \sinh^2{x} = 1$, we get
\begin{eqnarray}
    \mathcal{W}_{\phi^2}(\Delta\tau) = \frac{a^{4}}{128\pi^{4}}\sinh^{-4}{\left( \frac{\Delta\tau - 2i\epsilon}{2a^{-1}} \right)},
\end{eqnarray}
and using the relationship
\begin{eqnarray}
    \sinh^{-4}{x} = \sum^{\infty}_{k=-\infty} (x - i\pi k)^{-4} - \frac{2}{3} \sum^{\infty}_{k=-\infty} (x - i\pi k)^{-2},
\end{eqnarray}
where through this relationship we can finally write
\begin{eqnarray}
    \mathcal{W}_{\phi^2}(\Delta\tau) = \frac{1}{8\pi^{4}} \left[ \sum^{\infty}_{k=-\infty} (\Delta\tau - 2i\epsilon - 2\pi ik/a)^{-4} - \frac{a^2}{6}\sum^{\infty}_{k=-\infty} (\Delta\tau - 2i\epsilon - 2\pi ik/a)^{-2}\right].
    \label{func_Wigthman_final}
\end{eqnarray}
This expression represents the Wightman function for the quadratic coupled scalar field. Take note that it satisfies the KMS condition similarly to the linear coupling case.

\subsection{Transition probability rates}

In this section, we will look for an analytical form for the transition probability rate to a detector that interacts quadratically with the scalar field for a finite time. As we have seen in Sec. \ref{Sec IIA}, the transition probability rate depends, basically, on the response function of the detector, where the latter depends on the correlation function of vacuum. By employing the identity \cite{padmanabhan1982general}
\begin{eqnarray}
    f(u)[\e^{-i\Omega u}\mathcal{W}(u)] = f\bigg(-i\frac{\partial}{\partial \Omega}\bigg)[\e^{-i\Omega u}\mathcal{W}(u)],
\end{eqnarray}
valid for any function $f(u)$ that admits a power series expansion around $u = 0$, we can derive an asymptotic expression for the transition probability corresponding to any analytic window function as
\begin{eqnarray}
    \mathcal{F}^{\pm}_{\phi^2} = \chi\left(i\frac{\partial}{\partial\Omega}\right)\chi\left(-i\frac{\partial}{\partial\Omega}\right)\mathcal{F}^{\pm}_{\phi^2}(\infty), 
\end{eqnarray}
where $\mathcal{F}^{\pm}_{\phi^2}(\infty)$ corresponds to the infinite-time detector (for quadratic coupling)
\begin{eqnarray}
\mathcal{F}^{\pm}_{\phi^2}(\infty) = \int_{-\infty}^{\infty} \d\tau  \int_{-\infty}^{\infty} \d\tau' \e^{\pm i\Omega(\tau - \tau')}\mathcal{W}_{\phi^2}(\tau - \tau').
\end{eqnarray}
Expanding $\chi(\tau)$ as a Taylor series around $\tau = 0$ and assuming that $\chi(0) = 1$ and $\chi'(0) = 0$, we obtain that
\begin{eqnarray}
    \mathcal{F}^{\pm}_{\phi^2} \approx \mathcal{F}^{\pm}_{\phi^2}(\infty) - \chi''(0)\frac{\partial^2\mathcal{F}^{\pm}_{\phi^2}(\infty)}{\partial\Omega^2}.
\end{eqnarray}
Consequently, the corresponding transition probability rate is given by
\begin{eqnarray}
    \mathcal{R}^{\pm}_{\phi^2} \approx \mathcal{R}^{\pm}_{\phi^2}(\infty) - \chi''(0)\frac{\partial^2\mathcal{R}^{\pm}_{\phi^2}(\infty)}{\partial\Omega^2}
\end{eqnarray}
where
\begin{eqnarray}
    \mathcal{R}^{\pm}_{\phi^2}(\infty) = \int_{-\infty}^{\infty} \d(\Delta\tau) \e^{\pm i\Omega\Delta\tau}\mathcal{W}_{\phi^2}(\Delta\tau).
    \label{R_infty}
\end{eqnarray}

It is important to emphasize that the transition probability is sensitive to the derivatives of the switching (window) function. Therefore, an abrupt activation or deactivation of the detector can result in divergent contributions. To outsmart such divergences, we adopt a Gaussian window function given by $\chi(\tau) = \exp\left(-\frac{\tau^2}{2T^2}\right)$, which ensures a smooth switching on and off of the detector. Under this choice, and to leading order, the finite-time corrections to the transition probability rate can be expressed as
\begin{eqnarray}
    \mathcal{R}^{\pm}_{\phi^2} \approx \mathcal{R}^{\pm}_{\phi^2}(\infty) + \frac{1}{2T^2}\frac{\partial^2\mathcal{R}^{\pm}_{\phi^2}(\infty)}{\partial\Omega^2} + \mathcal{O}\left(T^{-4}\right).
    \label{R}
\end{eqnarray}
It is worth noting that the transition rate for a finite time of interaction $\mathcal{R}^{\pm}_{\phi^2}$ depends only on the transition rate that interacts during an infinite time $\mathcal{R}^{\pm}_{\phi^2}(\infty)$. In this way, substituting Eq.~(\ref{func_Wigthman_final}) into Eq.~(\ref{R_infty}), we have
\begin{eqnarray}
    \mathcal{R}^{-}_{\phi^2}(\infty) = \frac{1}{8\pi^{4}} \sum^{\infty}_{k=-\infty}\Bigg[ \int_{-\infty}^{\infty} \frac{\e^{- i\Omega\Delta\tau} \d(\Delta\tau)}{(\Delta\tau - 2i\epsilon - 2\pi ik/a)^{4}} - \frac{a^2}{6} \int_{-\infty}^{\infty} \frac{\e^{- i\Omega \Delta\tau}\d(\Delta\tau)}{(\Delta\tau - 2i\epsilon - 2\pi ik/a)^{2}}\Bigg],
    \label{R_infty2}
\end{eqnarray}
where we have for these two integrals the well-known solutions
\begin{eqnarray}
    \int_{-\infty}^{\infty} \frac{\e^{- i\Omega\Delta\tau} \d(\Delta\tau)}{(\Delta\tau - 2i\epsilon - 2\pi ik/a)^{2}} &=& 2\pi\Omega \exp{\left( \frac{2\pi\Omega k}{a}\right)}, \label{int1}\\
    \int_{-\infty}^{\infty} \frac{\e^{- i\Omega\Delta\tau} \d(\Delta\tau)}{(\Delta\tau - 2i\epsilon - 2\pi ik/a)^{4}} &=& -\frac{\pi}{3} \Omega^{3} \exp{\left( \frac{2\pi\Omega k}{a}\right)}.
    \label{int2}
\end{eqnarray}
It is worth emphasizing that the sum over $k$ is from $k=-\infty$ to $k=+\infty$, but the choice of contour for the integrals subtracts the contribution to the sum from $k \in [-\infty,0)$, and substituting Eq.~(\ref{int1}) and Eq.~(\ref{int2}) into Eq.~(\ref{R_infty2}), we obtain
\begin{eqnarray}
    \mathcal{R}^{-}_{\phi^2}(\infty) = -\frac{\Omega(a^2 + \Omega^2)}{24\pi^{3}} \sum^{\infty}_{k=0} \exp{\left( \frac{2\pi\Omega k}{a}\right)},
\end{eqnarray}
and consequently leading to a geometric series, i.e., $\sum^{\infty}_{k=0} \exp{\left(\frac{2\pi\Omega k}{a}\right)} = \left[1 - \exp{\left(\frac{2\pi\Omega}{a}\right)} \right]^{-1}$, and we get
\begin{eqnarray}
    \mathcal{R}^{-}_{\phi^2}(\infty) = \frac{(a^2 + \Omega^2)}{24\pi^{3}} \frac{\Omega}{\e^{2\pi\Omega/a} - 1},
    \label{R-_infty_final}
\end{eqnarray}
where this expression means the excitation probability rate for an infinite interaction time, for a quadratic coupling of the scalar field. Similarly, we have for the de-excitation probability rate,
\begin{eqnarray}
    \mathcal{R}^{+}_{\phi^2}(\infty) = \e^{2\pi\Omega/a}\mathcal{R}^{-}_{\phi^2}(\infty),
    \label{R+_infty_final}
\end{eqnarray}
for the conditions of quadratic coupling and infinite interaction time. Note that we can compare these functions with the case of linear coupling, i.e.,
\begin{eqnarray}
    \mathcal{R}^{\pm}_{\phi^2}(\infty) = \frac{(a^2 + \Omega^2)}{12\pi^{2}}\mathcal{R}_{\phi}^{\pm}(\infty),
    \label{relation_linear_quadratic}
\end{eqnarray}
where $\mathcal{R}_{\phi}^{\pm}(\infty)$ is the transition probability rate for an infinite interaction time, for a linear coupling of the scalar field.

Now, substituting Eq.~(\ref{R-_infty_final}) and Eq.~(\ref{R+_infty_final}) into Eq.~(\ref{R}), calculating the derivatives, and after several tedious algebraic manipulations, we obtain
\begin{eqnarray}
    \mathcal{R}^{-}_{\phi^2} &\approx& \frac{\Omega}{24\pi^{3}}\frac{(\Omega^2 + a^2)}{(\e^{2\pi\Omega/a} - 1)} \Bigg\{ 1 + \frac{3}{(\Omega^2 + a^2)T^2} +\nonumber\\
    &-&\frac{2\pi}{a\Omega T^2} \frac{\e^{2\pi\Omega/a}}{(\e^{2\pi\Omega/a} - 1)} \Bigg[ 
\frac{\pi\Omega}{a} \left( 1 - \frac{2\e^{2\pi\Omega/a}}{\e^{2\pi\Omega/a} -1}\right) + \frac{3\Omega^2 + a^2}{\Omega^2 + a^2} \Bigg]\Bigg\},
\label{R-}
\end{eqnarray}
and similarly,
\begin{eqnarray}
    \mathcal{R}^{+}_{\phi^2} &\approx& \frac{\Omega\e^{2\pi\Omega/a}}{24\pi^{3}}\frac{(\Omega^2 + a^2)}{(\e^{2\pi\Omega/a} - 1)} \Bigg\{ 1 + \frac{3}{(\Omega^2 + a^2)T^2} +\nonumber\\
    &-&\frac{2\pi}{a\Omega T^2} \frac{1}{(\e^{2\pi\Omega/a} - 1)} \Bigg[ 
    \frac{\pi\Omega}{a} \left( 1 - \frac{2\e^{2\pi\Omega/a}}{\e^{2\pi\Omega/a} -1}\right) + \frac{3\Omega^2 + a^2}{\Omega^2 + a^2} \Bigg]\Bigg\}.
\label{R+}
\end{eqnarray}
These expressions [Eq.~(\ref{R-}) and Eq.~(\ref{R+})] characterize the transition probability rates between the internal states of the detector that interacts (during a finite time) quadratically with the scalar field\footnote{For a comparison of both the linear and quadratic cases of transition probability rates, see the results for these rates for the linear case in the following references \cite{padmanabhan1982general, Pedro2024robustness}.}. It is worth noting that, at a finite interaction time, the detector does not have enough time to reach complete thermal equilibrium with the quantum field, resulting in a slight modification of the ideal thermal radiation \cite{Svaiter1992Inertial, Higuchi1993Uniformly, padmanabhan1982general}.



\section{Applications to quantum systems\label{sec:3}}

\subsection{Accelerated single--qubit}

\subsubsection{The physical model}

In quantum computing, the qubit (quantum bit) constitutes the fundamental unit of information \cite{nielsen2001quantum}, effectively replacing the classical bit employed in traditional computing architectures. Unlike its classical counterpart, a qubit can exist in a coherent superposition of its ground and excited states simultaneously, a property that underpins many of the advantages offered by quantum computation \cite{nielsen2001quantum, wilde2013quantum, khang2024applications}.

Now, we undertake an investigation of the influence of the Unruh effect on the quantum coherence of a uniformly accelerated single qubit. In particular, we aim to elucidate the impact of quadratic coupling on the fundamental properties of two-level systems. To this end, we consider a model comprising a detector interacting quadratically with a scalar quantum field. Within this framework, the field, situated in a Minkowski spacetime background, is initially prepared in the vacuum state $\vert 0_{\mathcal{M}}\rangle$, while the detector is initialized in a general qubit state, namely,
\begin{eqnarray}
|\psi_{\mathrm{D}}\rangle = \alpha|g\rangle + \beta|e\rangle,
\end{eqnarray}
where $\alpha$ and $\beta$ are complex amplitudes given by $\alpha = \e^{i\frac{\varphi}{2}}\cos{\frac{\theta}{2}}$, $\beta = \e^{-i\frac{\varphi}{2}}\sin{\frac{\theta}{2}}$, and they satisfy the relationship $\vert\alpha\vert^2 + \vert\beta\vert^2=1$.  Here, $\theta \in [0,\pi]$ and $\varphi \in [0,2\pi]$ are the polar and azimuthal angles of the Bloch sphere\footnote{The Bloch sphere provides a geometric representation that facilitates the visualization of quantum operations acting on a single qubit \cite{jazaeri2019review}. Each point on the surface of the sphere corresponds to a pure state of the qubit, thereby offering intuitive insight into its state evolution and coherence properties. For additional details, see ref.~\cite{kasirajan2021quantum}.}, respectively. Consequently, the density matrix $\hat{\rho}^{\mathrm{in}}_D = |\psi_D\rangle \langle\psi_D|$ corresponding to the detector quantum state $\vert\psi_D\rangle$, is written explicitly as
\begin{eqnarray}
    \hat{\rho}^{\mathrm{in}}_D = |\alpha|^2 |g\rangle\langle g| + \alpha\beta^*|g\rangle\langle e| + \alpha^*\beta|e\rangle\langle g| + |\beta|^2|e\rangle\langle e|.
    \label{rhoinD}
\end{eqnarray}

We consider a UDW detector initially prepared in a qubit state. This detector interacts quadratically with a massless quantum field $\phi$ over a finite time interval $T$. Following this interaction, it becomes possible to measure the internal states $\vert g\rangle$ and $\vert e\rangle$, thereby enabling the detection of modifications induced by the interaction. Mathematically, the interaction is characterized by $\hat{\rho}_{\mathrm{in}} = \hat{\rho}^{\mathrm{in}}_{D} \otimes \hat{\rho}_{\phi}$, where $\hat{\rho}_{\phi} = |0_{\mathcal{M}}\rangle \langle 0_{\mathcal{M}}|$.

The density matrix after the interaction is determined by the Hamiltonian describing the quadratic interaction [Eq.~(\ref{Hint})], and we have
\begin{eqnarray}
    \hat{\rho}^{\mathrm{out}}_{\phi^2} &=& \mathcal{\hat{U}}^{(0)}_{\phi^2} \hat{\rho}_{\mathrm{in}} \mathcal{\hat{U}}^{(0)^\dagger}_{\phi^2} + \mathcal{\hat{U}}^{(1)}_{\phi^2} \hat{\rho}_{\mathrm{in}} + \hat{\rho}_{\mathrm{in}}\mathcal{\hat{U}}^{(1)^\dagger}_{\phi^2} + \mathcal{\hat{U}}^{(1)}_{\phi^2} \hat{\rho}_{\mathrm{in}} \mathcal{\hat{U}}^{(1)^\dagger}_{\phi^2} + \mathcal{\hat{U}}^{(2)}_{\phi^2} \hat{\rho}_{\mathrm{in}} + \hat{\rho}_{\mathrm{in}} \mathcal{\hat{U}}^{(2)^\dagger}_{\phi^2} + \mathcal{O}(\lambda^3_{\phi^2}).
\end{eqnarray}
In this way, $\hat{\mathcal{U}}_{\phi^2}$ is the time evolution operator in perturbative form, which has the quadratic term of the scalar field, and is written explicitly as
\begin{eqnarray}
    \mathcal{\hat{U}}_{\phi^2} = \mathcal{\hat{U}}^{(0)}_{\phi^2} + \mathcal{\hat{U}}^{(1)}_{\phi^2} + \mathcal{\hat{U}}^{(2)}_{\phi^2} + \mathcal{O}(\lambda^3_{\phi^2}),
    \label{U}
\end{eqnarray}
with the following perturbative terms
\begin{eqnarray}
    \mathcal{\hat{U}}^{(0)}_{\phi^2} &=& \mathbb{I};
\end{eqnarray}
\begin{eqnarray}
    \mathcal{\hat{U}}^{(1)}_{\phi^2} &=& -i\lambda_{\phi^2}\int_{-\infty}^{\infty}d\tau\chi(\tau)\mu(\tau)\phi^{2}[x(\tau)];\\
    \mathcal{\hat{U}}^{(2)}_{\phi^2} &=& -\lambda^2_{\phi^2}\int^{+\infty}_{-\infty} d\tau \int^{+\tau}_{-\infty} d\tau' \chi(\tau)\chi(\tau')  \mu(\tau)\mu(\tau') \phi^{2}[x(\tau)]\phi^{2}[x(\tau')],
\end{eqnarray}
where $\mathbb{I}$ denotes the identity operator, and
\begin{eqnarray}
    \mu(\tau) = [\hat{\sigma}_{+}\e^{i\Omega\tau} + \hat{\sigma}_{-}\e^{-i\Omega\tau}]
    = \begin{pmatrix}
    0 & \e^{+i\Omega\tau} \\
    \e^{-i\Omega\tau} & 0
\end{pmatrix},
\end{eqnarray}
which promote transitions between the ground and excited states of the detector, and can be physically interpreted as the detector's response -- or click -- to the presence of the field. Furthermore, the operators $\hat{\sigma}_{+} = |e\rangle\langle g|$ and $\hat{\sigma}_{-} = |g\rangle\langle e|$ are defined as the creation and annihilation operators, respectively.

We now proceed to examine the final state of the UDW detector. In order to carry out this analysis, it is necessary to trace out the contributions associated with the field configuration. This procedure leads to
\begin{eqnarray}
    \hat{\rho}^{\mathrm{out}}_{D,\phi^2} = \mathrm{Tr}_{\vert0_{\mathcal{M}}\rangle}[\hat{\rho }^{\mathrm{out}}_{\phi^2}].
\end{eqnarray}
The reduced density matrix for the detector results in the following expression:
\begin{eqnarray}
    \hat{\rho}^{\mathrm{out}}_{D,\phi^2} = \hat{\rho}^{\mathrm{in}}_{D} + \mathrm{Tr}_{\vert0_{\mathcal{M}}\rangle} \left(\mathcal{\hat{U}}^{(1)}_{\phi^2} \hat{\rho}_{\mathrm{in}} \mathcal{\hat{U}}^{(1)\dagger}_{\phi^2}\right) + \mathrm{Tr}_{\vert0_{\mathcal{M}}\rangle} \left(\mathcal{\hat{U}}^{(2)}_{\phi^2}\hat{\rho}_{\mathrm{in}}\right) + \mathrm{Tr}_{\vert0_{\mathcal{M}}\rangle} \left(\hat{\rho}_{\mathrm{in}}\mathcal{\hat{U}}^{(2)\dagger}_{\phi^2}\right),
    \label{rhooutD}
\end{eqnarray}
and after performing extensive and meticulous calculations, we obtain that
\begin{eqnarray}
\mathrm{Tr}_{\vert0_{\mathcal{M}}\rangle} \left(\mathcal{\hat{U}}^{(1)}_{\phi^2} \hat{\rho}_{\mathrm{in}} \mathcal{\hat{U}}^{(1)\dagger}_{\phi^2}\right) &=& \lambda^2_{\phi^2} \int^{+\infty}_{-\infty} \d\tau \int^{+\infty}_{-\infty} \d\tau' \chi(\tau)\chi(\tau') \mathcal{W}_{\phi^2}(\tau, \tau') \nonumber\\
    &\times& \Big[\alpha\beta^* \e^{i\Omega(\tau+\tau')}|e\rangle\langle g| + |\alpha|^2 \e^{-i\Omega(\tau-\tau')}|e\rangle\langle e| + \nonumber\\
    &+& |\beta|^2 \e^{-i\Omega(\tau-\tau')}|g\rangle\langle g| + \beta\alpha^* \e^{-i\Omega(\tau+\tau')}|g\rangle\langle e|\Big],\label{term1}
\end{eqnarray}
\begin{eqnarray}
\mathrm{Tr}_{\vert0_{\mathcal{M}}\rangle} \left(\mathcal{\hat{U}}^{(2)}_{\phi^2}\hat{\rho}_{\mathrm{in}}\right) &=& -\lambda^2_{\phi^2} \int^{+\infty}_{-\infty} \d\tau \int^{+\tau}_{-\infty} \d\tau_1 \chi(\tau)\chi(\tau_1) \mathcal{W}_{\phi^2}(\tau, \tau_1) \nonumber\\
    &\times& \Big[\e^{i\Omega(\tau-\tau_1)} \left( \beta\alpha^*|e\rangle\langle g| + |\beta|^2|e\rangle\langle e| \right) + \nonumber\\
    &+& \e^{-i\Omega(\tau-\tau_1)} \left( |\alpha|^2|g\rangle\langle g| + \alpha\beta^*|g\rangle\langle e| \right) \Big],
    \label{term2}
\end{eqnarray}
\begin{eqnarray}
\mathrm{Tr}_{\vert0_{\mathcal{M}}\rangle} \left(\hat{\rho}_{\mathrm{in}}\mathcal{\hat{U}}^{(2)\dagger}_{\phi^2}\right) &=& -\lambda^2_{\phi^2} \int^{+\infty}_{-\infty} \d\tau \int^{+\tau}_{-\infty} \d\tau_1 \chi(\tau)\chi(\tau_1) \mathcal{W}^{*}_{\phi^2}(\tau, \tau_1) \nonumber\\
    &\times& \Big[\e^{-i\Omega(\tau-\tau_1)} \left( \alpha\beta^*|g\rangle\langle e| + |\beta|^2|e\rangle\langle e| \right) + \nonumber\\
    &+& \e^{i\Omega(\tau-\tau_1)} \left( |\alpha|^2|g\rangle\langle g| + \beta\alpha^*|e\rangle\langle g| \right)\Big].
    \label{term3}
\end{eqnarray}

Therefore, in this conjecture, substituting Eq.~(\ref{term1}), Eq.~(\ref{term2}), and Eq.~(\ref{term3}) in Eq.~(\ref{rhooutD}), and defining the integrals as follows
\begin{eqnarray}
    \mathcal{C}^{\pm}_{\phi^2} &=& \int^{+\infty}_{-\infty} \d\tau \int^{+\infty}_{-\infty} \d\tau' \chi(\tau)\chi(\tau') \e^{\pm i\Omega(\tau+\tau')} \mathcal{W}_{\phi^2}(\tau, \tau'),
    \label{C+-}
\end{eqnarray}
\begin{eqnarray}
    \mathcal{G}^{\pm}_{\phi^2} &=& \int^{+\infty}_{-\infty} \d\tau \int^{+\tau}_{-\infty} \d\tau' \chi(\tau)\chi(\tau')  \e^{\pm i\Omega(\tau-\tau')} \mathcal{W}_{\phi^2}(\tau, \tau'),
    \label{intG}
\end{eqnarray}
\begin{eqnarray}
    \mathcal{F}^{\pm}_{\phi^2} &=& \int_{-\infty}^{+\infty} \d\tau \int_{-\infty}^{+\infty} \d\tau' \chi(\tau)\chi(\tau') \e^{\pm i\Omega(\tau - \tau')} \mathcal{W}_{\phi^2}(\tau, \tau'),
    \label{Flw}
\end{eqnarray}
and omitting from now on the superscript ``$\mathrm{out}$'' to simplify notation, finally, one obtains
\begin{eqnarray}
    \hat{\rho}_{D,\phi^{2}} &=& \left\{ \cos^2{\frac{\theta}{2}} + \lambda^{2}_{\phi^2}\left[ \sin^2{\frac{\theta}{2}} \mathcal{F}^{-}_{\phi^2} - 2\cos^2{\frac{\theta}{2}} \mathbf{Re}\left(\mathcal{G}^{-}_{\phi^2}\right) \right] \right\} |g\rangle \langle g| \nonumber\\
    &+& \left\{ \sin^2{\frac{\theta}{2}} + \lambda^{2}_{\phi^2}\left[ \cos^2{\frac{\theta}{2}} \mathcal{F}^{+}_{\phi^2} - 2\sin^2{\frac{\theta}{2}} \mathbf{Re}\left(\mathcal{G}^{+}_{\phi^2}\right) \right] \right\} |e\rangle \langle e| \nonumber\\
    &+& \left\{ \frac{1}{2}\e^{-i\varphi}\sin{\theta} + \frac{\lambda^{2}_{\phi^2}}{2} \left[\e^{i\varphi}\mathcal{C}^{+}_{\phi^2} - \e^{-i\varphi} \left(\mathcal{G}^{+}_{\phi^2} + \mathcal{G}^{-*}_{\phi^2}\right) \right]\sin{\theta} \right\} |e\rangle \langle g| \nonumber\\
    &+& \left\{ \frac{1}{2}\e^{+i\varphi}\sin{\theta} + \frac{\lambda^{2}_{\phi^2}}{2}  \left[\e^{-i\varphi}\mathcal{C}^{-}_{\phi^2} - \e^{i\varphi} \left(\mathcal{G}^{-}_{\phi^2} + \mathcal{G}^{+*}_{\phi^2}\right) \right] \sin{\theta} \right\} |g\rangle \langle e|.
    \label{final rho}
\end{eqnarray}
It is noteworthy that the reduced density matrix Eq.~(\ref{final rho}) is traceless in terms proportional to $\lambda_{\phi^2}^2$. Here in this setup $\mathbf{Re}(\mathcal{G}^-) = \mathbf{Re}(\mathcal{G}^+)$, thus note that,
\begin{eqnarray}
    \mathbf{Re}(\mathcal{G}^{-}_{\phi^2}) = \frac{1}{2} \left( \mathcal{F}^{-}_{\phi^2}\sin^2{\frac{\theta}{2}} + \mathcal{F}^{+}_{\phi^2} \cos^2{\frac{\theta}{2}}\right).
    \label{RelationTraceless}
\end{eqnarray}
Additionally, we have explicitly obtained an analytical solution for the integral $\mathcal{C}^{\pm}_{\phi^2}$, see Appendix \ref{AppendixIntegralC} for details. Therefore, this integral has the following solution
\begin{eqnarray}
    \mathcal{C}^\pm_{\phi^2} &=& \frac{aT\sqrt{\pi}}{48\pi^4} \left( a^2 + \frac{1}{2T^2} \right)\e^{-T^2\Omega^2}.
    \label{Solution_Integral_C}
\end{eqnarray}
On the other hand, this expression decays exponentially as a function of the interaction time of the detector-field system.

\subsubsection{Quantum coherence}

It is worth mentioning that we cannot ``see'' a superposition in itself, but only classical states \cite{griffiths2018introduction}. We cannot determine in advance what the outcome of the measurement will be. The only thing we can say before the measurement is that we will observe a specific state with a corresponding probability \cite{sakurai2020modern}. On the other hand, we can obtain information about the superposition of a system through the well-known quantum coherence.

Quantum coherence is a fundamental property characterized by the presence of quantum superposition, which permits interference between distinct eigenstates \cite{Leggett1980}. Furthermore, it is defined by the preservation of relative phases between the components of a superposed quantum state, thereby allowing essential quantum phenomena such as interference and entanglement \cite{streltsov2015measuring}. Within this framework, quantum optical techniques constitute a vital set of tools for the manipulation and control of coherence \cite{Glauber1963coherent,sudarshan1963}.

In the specific case of a two-level quantum system, the coherence between the states $\vert g\rangle$ and $\vert e\rangle$ is quantified using the $l^1$ norm quantum coherence, defined as the sum of the absolute values of the off-diagonal elements of the system's density matrix \cite{Baumgratz2014QuantifyingCoherence}, as follows:
\begin{align}
    \mathcal{Q}^{l^1}_{\phi^2}(\hat{\rho}_{D,\phi^2}) = \sum_{i \neq j} \mid \hat{\rho}_{D,\phi^2}^{ij}\mid.
    \label{DefCoherence}
\end{align}
Developing Eq.~(\ref{DefCoherence}) considering $i$ and $j$ by permuting between the values of $g$ and $e$, we have
\begin{align}
    \mathcal{Q}^{l^1}_{\phi^2}(\hat{\rho}_{D,\phi^2}) = \mid \hat{\rho}_{D,\phi^2}^{ge}\mid + \mid \hat{\rho}_{D,\phi^2}^{ge}\mid = 2 \sqrt{(\hat{\rho}_{D,\phi^2}^{ge})(\hat{\rho}_{D,\phi^2}^{ge})^{*}},
    \label{Cl1}
\end{align}
where $(\hat{\rho}_{D,\phi^2}^{ge})^{*} = \hat{\rho}_{D,\phi^2}^{eg}$ and
\begin{eqnarray}
    \hat{\rho}^{\mathrm{ge}}_{D,\phi^2} &=& \frac{1}{2}\e^{+i\varphi}\sin{\theta} +  \frac{\lambda^{2}_{\phi^2}}{2} \left[\e^{-i\varphi}\mathcal{C}^{-}_{\phi^2} - \e^{i\varphi} \left(\mathcal{G}^{-}_{\phi^2} + \mathcal{G}^{+*}_{\phi^2}\right) \right]\sin{\theta} .
    \label{rev_n1}
\end{eqnarray}
Substituting Eq.~(\ref{rev_n1}) into Eq.~(\ref{Cl1}), we obtain
\begin{align}
    \mathcal{Q}^{l^1}_{\phi^2} = \vert \sin{\theta}\vert \left\{ 1 - \lambda_{\phi^2} \left[ 2\mathbf{Re}(\mathcal{G}^{-}_{\phi^2}) - \mathcal{C}^{-}_{\phi^2} \cos{(2\varphi)}\right]\right\} + \mathcal{O}(\lambda_{\phi^2}^{4}).
\end{align}

Now, using the relation given by Eq.~(\ref {RelationTraceless}), defining the following dimensionless parameters: $\Lambda = \lambda_{\phi^{2}}\Omega$, $\overline{a} = a/\Omega$, $\sigma = \Omega T$, and $\overline{\mathcal{R}}^{\pm}_{\phi^2} = \mathcal{R}^{\pm}_{\phi^2}/\Omega$; and using the Eq.~(\ref{Solution_Integral_C}), one obtains
\begin{eqnarray}
    \mathcal{Q}^{l^1}_{\phi^2} = \vert \sin{\theta}\vert \left\{ 1 - \sigma\Lambda^{2} \left[ \frac{\overline{\mathcal{R}}^{-}_{\phi^2}}{\Omega^2} \sin^{2}{\frac{\theta}{2}} + \frac{\overline{\mathcal{R}}^{+}_{\phi^2}}{\Omega^2} \cos^{2}{\frac{\theta}{2}} - \frac{\overline{a}\sqrt{\pi}}{48\pi^{4}} \left( 
\overline{a}^2 + \frac{1}{2\sigma^{2}}\right) \e^{-\sigma^2}\cos{(2\varphi)}\right]\right\},
\label{Coherence_2}
\end{eqnarray}
where the $\mathcal{F}^{\pm}_{\phi^{2}} = \sigma\overline{\mathcal{R}}^{\pm}_{\phi^2}$ was used. By substituting into Eq.~(\ref{Coherence_2}) the expressions for the excitation probability rate [Eq.~(\ref{R-})] and the de-excitation probability rate [Eq.~(\ref{R+})], and considering the regime of long interaction time, i.e., $\sigma \gg 1$, we get
\begin{eqnarray}
    \mathcal{Q}^{l^1}_{\phi^2} = \vert \sin{\theta}\vert \left[ 1 - \frac{(1 + \overline{a}^2)}{24\pi^{3}}\frac{\sigma\Lambda^{2}}{\e^{2\pi/\overline{a}} - 1} \left( \sin^{2}{\frac{\theta}{2}} + \e^{2\pi/\overline{a}}\cos^{2}{\frac{\theta}{2}}\right)\right] + \mathcal{O}(\Lambda^{4}),
\label{coherenceQubit_quadratic}
\end{eqnarray}
where the long interaction time limit has been taken to ensure that the detector has enough time to interact with the quantum field.
Additionally, it is important to be aware that if we take the limit for the massless field in Eq.~(3.21) of ref. \cite{pedro2025mitigating}, we will obtain the $l^1$ norm quantum coherence for a uniformly accelerated qubit linearly coupled with a scalar field, given as follows
\begin{eqnarray}
    \mathcal{Q}^{l^1}_{\phi} = \vert \sin{\theta}\vert \left[ 1 - \frac{1}{2\pi}\frac{\sigma\lambda^{2}}{\e^{2\pi/\overline{a}} - 1} \left( \sin^{2}{\frac{\theta}{2}} + \e^{2\pi/\overline{a}}\cos^{2}{\frac{\theta}{2}}\right)\right] + \mathcal{O}(\lambda^{4}).
\label{coherenceQubit_linear}
\end{eqnarray}

It is worth emphasizing that the main change from the linear case to the quadratic case is in the coupling constant, in addition, the second term on the right side of Eq.~(\ref{coherenceQubit_quadratic}) depends quadratically on the parameter $\overline{a}$ which is responsible for the acceleration. This in effect implies that the coherence will decay quadratically as a function of the acceleration.

\subsubsection{Probability of internal states}

In this section, we investigate the degradation of the probability amplitudes of internal states due to the Unruh effect. Thus, the probabilities associated with these states are defined using the reduced density matrix with respect to the detector [Eq.~(\ref{final rho})]. This formulation enables the investigation of the effects of the Unruh phenomenon on the probability amplitudes of the detector. For this purpose, the probability of finding the qubit in the ground state and in the excited state is defined, respectively, as follows:
\begin{eqnarray}
    P^{g}_{\phi^2} &=& \langle g|\hat{\rho}_{D,\phi^2}|g\rangle, \\
    P^{e}_{\phi^2} &=& \langle e|\hat{\rho}_{D,\phi^2}|e\rangle,
\end{eqnarray}
and this gives us
\begin{eqnarray}
    P^{g}_{\phi^{2}} &=& \cos^2{\frac{\theta}{2}} + \lambda^{2}_{\phi^2}\left[ \mathcal{F}^-_{\phi^2}\sin^2{\frac{\theta}{2}} - 2\mathbf{Re}\left(\mathcal{G}^-_{\phi^2}\right)\cos^2{\frac{\theta}{2}} \right], \\
    P^{e}_{\phi^{2}} &=& \sin^2{\frac{\theta}{2}} + \lambda^{2}_{\phi^2}\left[ \mathcal{F}^+_{\phi^2}\cos^2{\frac{\theta}{2}} - 2\mathbf{Re}\left(\mathcal{G}^+_{\phi^2}\right)\sin^2{\frac{\theta}{2}} \right].
\end{eqnarray}
Using the relation $\mathcal{F}^{\pm}_{\phi^{2}} = \sigma\overline{\mathcal{R}}^{\pm}_{\phi^2}$, substituting Eq. (\ref{Solution_Integral_C}), performing some algebraic operations, and applying the limit $\sigma \gg 1$, we find the following expressions
\begin{eqnarray}
    P^{g}_{\phi^{2}} &=& \cos^2{\frac{\theta}{2}} + \frac{(1+\overline{a}^2)}{24\pi^3}\frac{\sigma\Lambda^{2}}{\e^{2\pi/\overline{a}} - 1} \left( \sin^4{\frac{\theta}{2}} - \e^{2\pi/\overline{a}} \cos^4{\frac{\theta}{2}} \right),
    \label{Pg_quadratic}\\
    P^{e}_{\phi^{2}} &=& \sin^2{\frac{\theta}{2}} - \frac{(1+\overline{a}^2)}{24\pi^3}\frac{\sigma\Lambda^{2}}{\e^{2\pi/\overline{a}} - 1} \left( \sin^4{\frac{\theta}{2}} - \e^{2\pi/\overline{a}} \cos^4{\frac{\theta}{2}} \right).
    \label{Pe_quadratic}
\end{eqnarray}
Note that the total probability $P^{\mathrm{total}}_{\phi^2} = P^{g}_{\phi^{2}} + P^{e}_{\phi^{2}} = 1$. Additionally, be aware that for the linear case we have
\begin{eqnarray}
    P^{g}_{\phi} &=& \cos^2{\frac{\theta}{2}} + \frac{1}{2\pi}\frac{\sigma\lambda^{2}}{\e^{2\pi/\overline{a}} - 1} \left( \sin^4{\frac{\theta}{2}} - \e^{2\pi/\overline{a}} \cos^4{\frac{\theta}{2}} \right),
    \label{Pg_linear}\\
    P^{e}_{\phi} &=& \sin^2{\frac{\theta}{2}} - \frac{1}{2\pi}\frac{\sigma\lambda^{2}}{\e^{2\pi/\overline{a}} - 1} \left( \sin^4{\frac{\theta}{2}} - \e^{2\pi/\overline{a}} \cos^4{\frac{\theta}{2}} \right),
    \label{Pe_linear}
\end{eqnarray}
these expressions can be obtained by applying the limit for the massless field in Eq.~(3.25) of ref. \cite{pedro2025mitigating}. Note that the change from the linear to the quadratic case is just the coupling constant and the factor $(1+\overline{a}^2)/12\pi^2$. Showing that for quadratic coupling the degradation of the probability amplitudes has a quadratic dependence as a function of the acceleration.

\subsubsection{Numerical results}

As seen in previous sections, quantum coherence and the probabilities of internal states for quadratic coupling behave differently when compared to linearly coupled cases. Regarding quantum coherence, when comparing Eq.~(\ref{coherenceQubit_quadratic}) with Eq.~(\ref{coherenceQubit_linear}), and regarding probabilities, when comparing Eq.~(\ref{Pg_quadratic}) and Eq.~(\ref{Pe_quadratic}) with Eq.~(\ref{Pg_linear}) and Eq.~(\ref{Pe_linear}), it is possible to notice that the difference between these expressions exists only in the second term on the right-hand side of each of them. This difference is given by the coupling constant, which for the linear coupling is given by $\lambda$ and for the quadratic coupling is given by $\Lambda$. Furthermore, for the quadratic case there is an additional multiplicative factor given by $(1+\overline{a}^2)/12\pi^2$.

\begin{figure}
    \centering
    \includegraphics[width=0.475\linewidth]{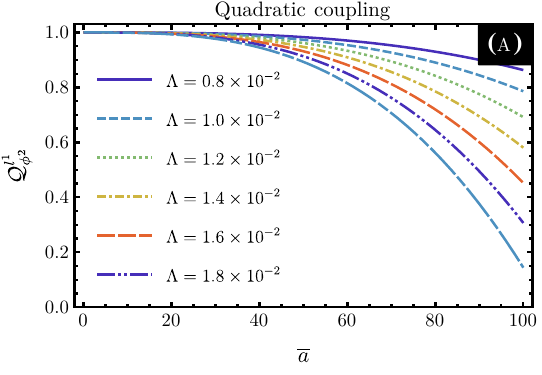}
    \includegraphics[width=0.49\linewidth]{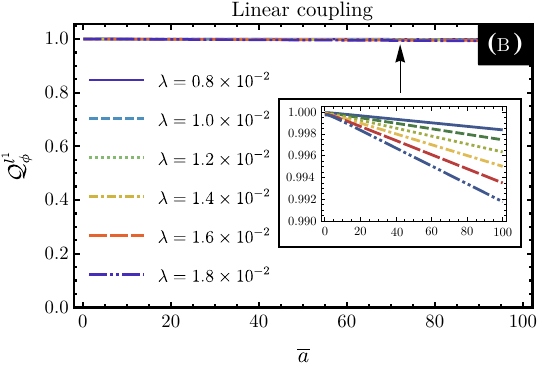}
    \includegraphics[width=0.475\linewidth]{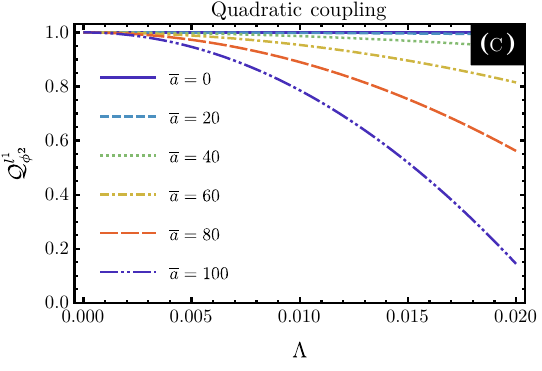}
    \includegraphics[width=0.49\linewidth]{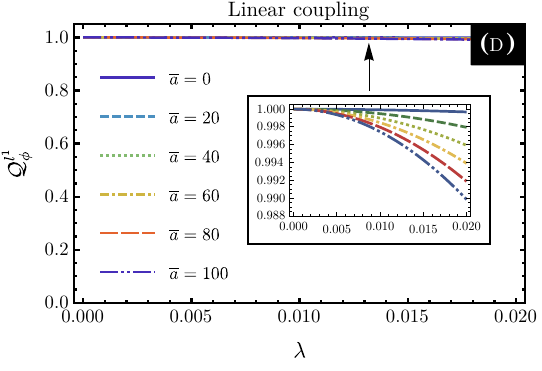}
    \includegraphics[width=0.49\linewidth]{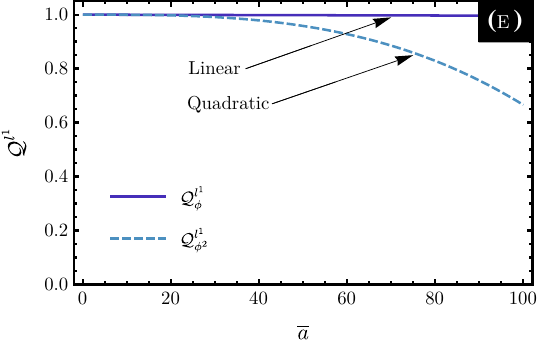}
    \caption{Behavior of $\mathcal{Q}^{l^1}$ as a function of the parameter $\overline{a}$: \textbf{(a)} for quadratic coupling with different values of $\Lambda$, and \textbf{(b)} for linear coupling with different values of $\lambda$. Dependence of $\mathcal{Q}^{l^1}$ on \textbf{(c)} $\Lambda$ and \textbf{(d)} $\lambda$ for different values of $\overline{a}$. \textbf{(e)} The comparison between $\mathcal{Q}^{l^1}_{\phi^2}$ and $\mathcal{Q}^{l^1}_{\phi}$ as functions of $\overline{a}$ where $\Omega = 1$ is implicitly assumed. In all cases, the parameters $\theta = \pi/2$ and $\sigma = 10$ are held constant.}
    \label{Fig1}
\end{figure}

In this way, observe Fig.~\ref{Fig1}(a) and Fig.~\ref{Fig1}(b), where we have the comparison of quantum coherence for the linear and quadratic cases, both as a function of $\overline{a}$. Note that for quadratic coupling we have a much larger degradation of coherence when compared to the linear coupling case.
In Fig. \ref{Fig1}(c) and Fig. \ref{Fig1}(d), we have the comparison of quantum coherence for the two types of couplings, each one as a function of its respective coupling constant. Note that, considering the range chosen for the coupling constants, the quadratic case degrades the coherence quite considerably. Also note that for the linear case it is necessary to zoom in on the plot to see the degradation of coherence significantly. 

Aiming at a more direct visual comparison, we plot Fig.~\ref{Fig1}(e). This plot shows the behavior of coherence for the two types of couplings, both as a function of the parameter $\overline{a}$, where implicitly assuming that $\Omega = 1$ we can adopt that $\lambda = \Lambda = 1.25 \times 10^{-2}$. Note that, for these conditions, quantum coherence for the quadratic coupling case decreases rapidly, unlike the linear case, which becomes imperceptible to its degradation. This indicates that the loss of coherence happens much faster when the detector is quadratically coupled with the scalar field. This highlights the difference in strength between the two types of couplings in question.

\begin{figure}[htb]
    \centering
    \includegraphics[width=0.67\linewidth]{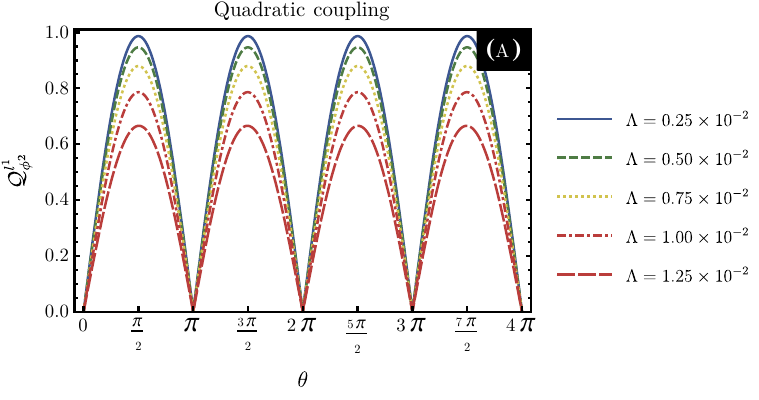}
    \includegraphics[width=0.67\linewidth]{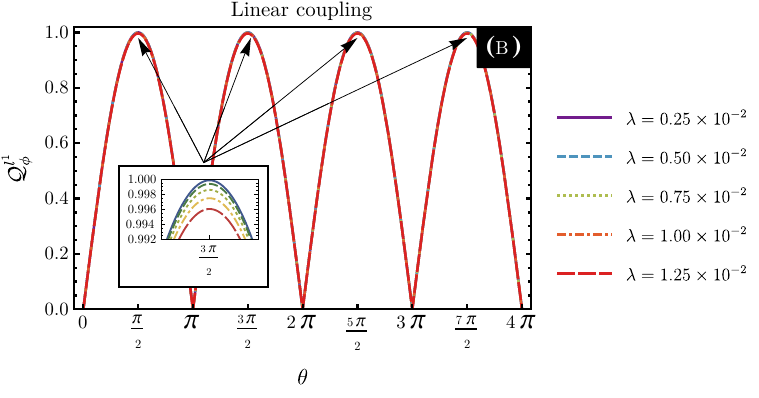}
    \includegraphics[width=0.67\linewidth]{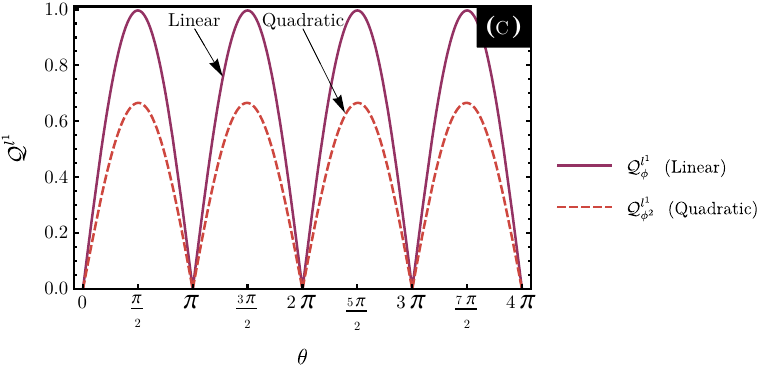}
    \caption{$\mathcal{Q}^{l^1}$ as a function of the polar angle $\theta$: \textbf{(a)} for quadratic coupling with different values of $\Lambda$ and \textbf{(b)} for linear coupling with different values of $\lambda$. \textbf{(c)} Plot of $\mathcal{Q}^{l^1}_{\phi^2}$ and $\mathcal{Q}^{l^1}_{\phi}$ as functions of the $\theta$ (implicitly assuming that $\Omega = 1$). The parameters $\overline{a} = 100$ and $\sigma = 10$ are held constant.}
    \label{Fig2}
\end{figure}

In Fig. \ref{Fig2}(a) and Fig. \ref{Fig2}(b), we compare the oscillatory behavior\footnote{Note that, since coherence quantifies the system’s ability to remain in superposition and the qubit’s polar angle determines this superposition, analyzing coherence as a function of the polar angle results in an oscillatory profile due to the mixing of internal states.} of the coherence in terms of the polar angle $\theta$ for different values of their respective coupling constants.
The coherence amplitude reaches its maximum when $\theta$ is a half-integer multiple of $\pi$, indicating the presence of maximal superposition between mixed states. In contrast, the coherence amplitude vanishes when $\theta$ is an integer multiple of $\pi$, corresponding to the poles of the Bloch sphere where the qubit resides in a definite state with no superposition. Additionally, it is observed that if $\Lambda$ or $\lambda$ increases, the amplitude of the oscillation will decrease, suggesting an amplification in the coherence degradation, this amplification being much larger for the quadratic case compared to the linear case.

For a more direct visual comparison, we plot Fig.~\ref{Fig2}(c). This plot shows the behavior of coherence for the two types of coupling, both as a function of the qubit polar angle, where we adopt $\lambda = \Lambda = 1.25 \times 10^{-2}$ (implicitly assuming that $\Omega = 1$). Note that for these conditions, the quantum coherence for the quadratic coupling case rapidly decreases the oscillation amplitude, unlike the linear case, which becomes imperceptible to its oscillation amplitude degradation.

\begin{figure}
    \centering
    \includegraphics[width=0.49\linewidth]{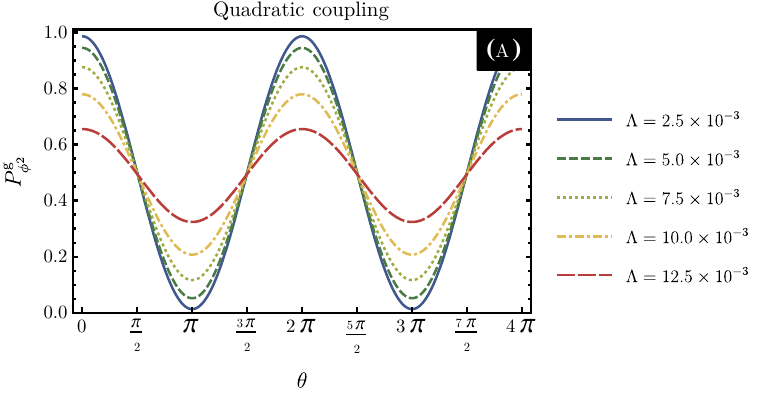}
    \includegraphics[width=0.49\linewidth]{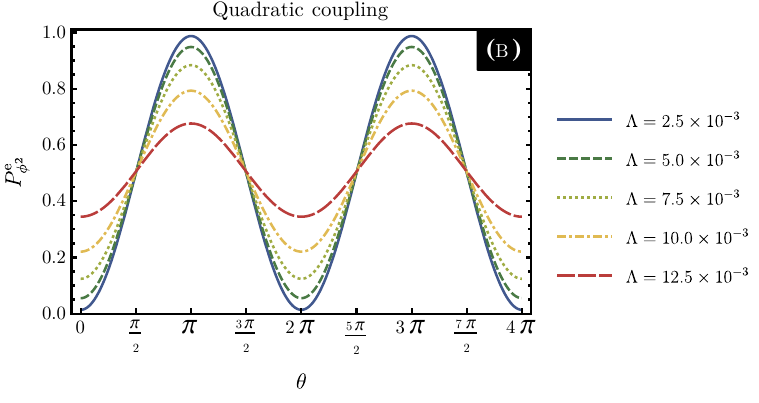}
    \includegraphics[width=0.49\linewidth]{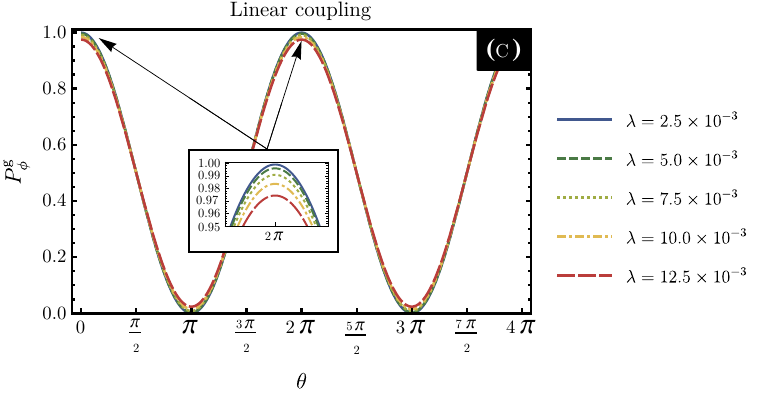}
    \includegraphics[width=0.49\linewidth]{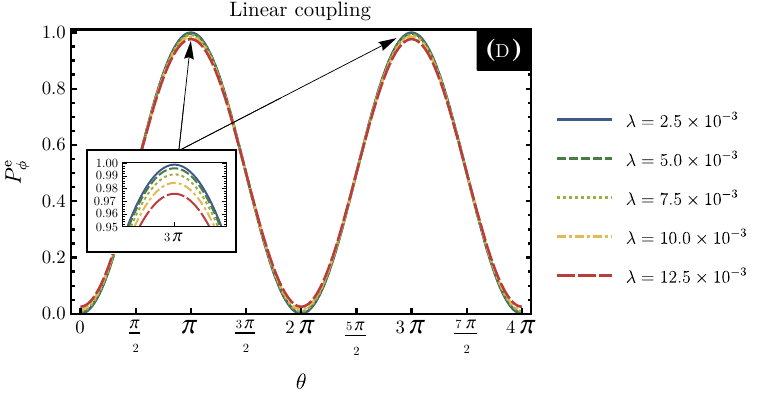}
    \includegraphics[width=0.49\linewidth]{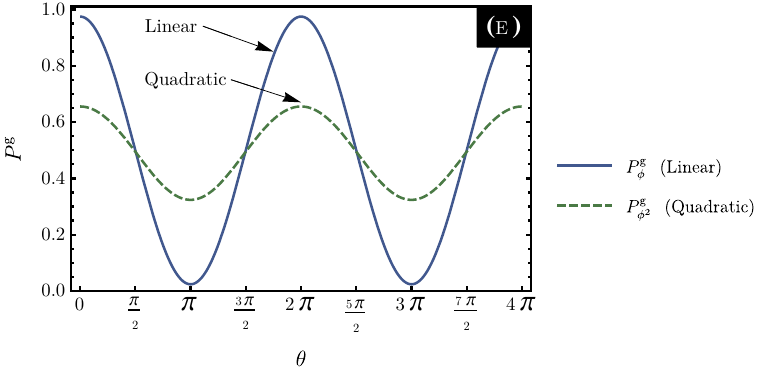}
    \includegraphics[width=0.49\linewidth]{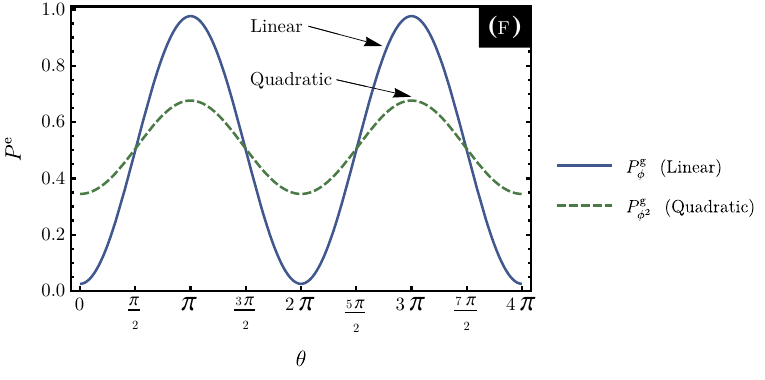}
    \caption{Probabilities \textbf{(a)} $P^{g}_{\phi^2}$ and \textbf{(b)} $P^{e}_{\phi^2}$ as functions of the polar angle $\theta$ for different values of $\Lambda$. Plot of the probabilities \textbf{(c)} $P^{g}_{\phi}$ and \textbf{(d)} $P^{e}_{\phi}$ as functions of the polar angle $\theta$ for different values of $\lambda$. Plots of \textbf{(e)} $P^{g}_{\phi}$ and $P^{g}_{\phi^2}$, and \textbf{(f)} $P^{e}_{\phi}$ and $P^{e}_{\phi^2}$ as functions of $\theta$ where it was implicitly assumed that $\Omega = 1$. The parameters $\overline{a} = 100$ and $\sigma = 10$ are held constant.}
    \label{Fig3}
\end{figure}

We also plot the ground state probability [Fig.~\ref{Fig3}(a)] and the excited state probability [Fig.~\ref{Fig3}(b)] for quadratic coupling as a function of the $\theta$ for different values of $\Lambda$. Note that the corresponding oscillation amplitudes degrade as $\Lambda$ increases. For the linear case, we plot the ground state probability [Fig.~\ref{Fig3}(c)] and the excited state probability [Fig.~\ref{Fig3}(d)] as a function of the $\theta$ for different values of $\lambda$. In this case, note that the degradation of the corresponding amplitudes is almost imperceptible, and it is necessary to zoom in to see significant effects.

For a more direct visual comparison, we present Fig.~\ref{Fig3}(e) and Fig.~\ref{Fig3}(f), which illustrate the behavior of the internal state probabilities for both types of coupling as functions of the parameter $\theta$. In these plots, we adopt $\lambda = \Lambda = 1.25 \times 10^{-2}$ and for this comparison it is implicitly assumed that $\Omega = 1$. It is evident that, under these conditions, the probabilities associated with the internal states in the case of quadratic coupling exhibit a more rapid decay in oscillation amplitude. In contrast, the linear coupling case demonstrates a significantly slower degradation, with the reduction in oscillation amplitude being almost imperceptible.

\subsection{Quantum interferometric circuits}

\subsubsection{The physical model}

In this section, we analyze the quantum scattering circuit. The circuit is organized as follows: initially, a single qubit is prepared in a well-defined initial state. Subsequently, the application of the first Hadamard gate transforms the qubit into a superposition state. Afterward, a phase accumulation occurs due to the phase-change gate $\hat{\alpha}$. Next, a unitary operator $\hat{\mathcal{U}}$ [Eq.~(\ref{U})] induces an quadratic interaction (in a controlled manner) between the detector and the quantum field. Finally, the qubit undergoes a second Hadamard operation, after which measurements on the system are performed.

In this setup, the qubit employed in the quantum interferometric circuit is initially represented by the density matrix $\hat{\rho}^{\mathrm{g}}_{D,I} = |g\rangle \langle g|$. The application of the first Hadamard gate given by
\begin{eqnarray}
    \hat{H} = \frac{1}{\sqrt{2}} \renewcommand{\arraystretch}{0.8}\begin{pmatrix}
    1 & 1 \\
    1 & -1
\end{pmatrix},
\end{eqnarray}
transforms the state into the following expression:
\begin{eqnarray}
    \hat{\rho}^{\mathrm{in}}_{D,I} = \frac{1}{2}\left( |g\rangle \langle g| + |g\rangle \langle e| + |e\rangle \langle g| + |e\rangle \langle e|\right).
\end{eqnarray}

During the evolution between the two Hadamard gates, the probability amplitudes for finding the qubit in the internal states undergo phase accumulation. Specifically, a phase given by
\begin{eqnarray}
    \hat{\alpha} = \frac{1}{\sqrt{2}} \renewcommand{\arraystretch}{0.8}\begin{pmatrix}
    1 & 0 \\
    0 & \e^{i\alpha}
\end{pmatrix},
\end{eqnarray}
is introduced by the phase-shift gate when the qubit is in the excited state $|e\rangle$. In this way, the initial density matrix of the detector becomes
\begin{eqnarray}
    \hat{\rho}^{\mathrm{in}}_{D,I} = \frac{1}{2}\left( |g\rangle \langle g| + \e^{-i\alpha}|g\rangle \langle e| + \e^{i\alpha}|e\rangle \langle g| + |e\rangle \langle e|\right).
\end{eqnarray}

Consequently, the initial state of the combined system is represented as $\hat{\rho}_{\mathrm{in},I} = \hat{\rho}^{\mathrm{in}}_I \otimes \hat{\rho}_\phi$. Thus, the density matrix after the interaction is given by
\begin{eqnarray}
    \hat{\rho}^{\mathrm{out}}_{\phi^2} &=& \mathcal{\hat{U}}^{(0)}_{\phi^2} \hat{\rho}_{\mathrm{in},I} \mathcal{\hat{U}}^{(0)^\dagger}_{\phi^2} + \mathcal{\hat{U}}^{(1)}_{\phi^2} \hat{\rho}_{\mathrm{in},I} + \hat{\rho}_{\mathrm{in},I}\mathcal{\hat{U}}^{(1)^\dagger}_{\phi^2} + \mathcal{\hat{U}}^{(1)}_{\phi^2} \hat{\rho}_{\mathrm{in},I} \mathcal{\hat{U}}^{(1)^\dagger}_{\phi^2} + \mathcal{\hat{U}}^{(2)}_{\phi^2} \hat{\rho}_{\mathrm{in},I} + \hat{\rho}_{\mathrm{in},I} \mathcal{\hat{U}}^{(2)^\dagger}_{\phi^2} + \mathcal{O}(\lambda^3_{\phi^2}).\nonumber\\
\end{eqnarray}
and applying the trace out to the degrees of freedom of the field, we obtain 
\begin{eqnarray}
    \hat{\rho}^{\mathrm{out}}_{I,\phi^2} = \hat{\rho}^{\mathrm{in}}_{D,I} + \mathrm{Tr}_{\vert0_{\mathcal{M}}\rangle} \left(\mathcal{\hat{U}}^{(1)}_{\phi^2} \hat{\rho}_{\mathrm{in},I} \mathcal{\hat{U}}^{(1)\dagger}_{\phi^2}\right) + \mathrm{Tr}_{\vert0_{\mathcal{M}}\rangle} \left(\mathcal{\hat{U}}^{(2)}_{\phi^2}\hat{\rho}_{\mathrm{in},I}\right) + \mathrm{Tr}_{\vert0_{\mathcal{M}}\rangle} \left(\hat{\rho}_{\mathrm{in},I}\mathcal{\hat{U}}^{(2)\dagger}_{\phi^2}\right),
    \label{rhooutI}
\end{eqnarray}
and after many algebraic manipulations, we get
\begin{eqnarray}
    \hat{\rho}_{I,\phi^{2}}^{\mathrm{out}} &=& \frac{1}{2}\left[ 1 + \lambda^{2}_{\phi^2}\left( \mathcal{F}^{-}_{\phi^2} - 2 \mathbf{Re}\left(\mathcal{G}^{-}_{\phi^2}\right) \right) \right] |g\rangle \langle g| \nonumber\\
    &+& \frac{1}{2}\left[ 1 + \lambda^{2}_{\phi^2}\left( \mathcal{F}^{+}_{\phi^2} - 2 \mathbf{Re}\left(\mathcal{G}^{+}_{\phi^2}\right) \right) \right] |e\rangle \langle e| \nonumber\\
    &+& \frac{1}{2}\left\{ \e^{-i\alpha} + \lambda^{2}_{\phi^2} \left[\e^{i\alpha}\mathcal{C}^{-}_{\phi^2} - \e^{-i\alpha} \left(\mathcal{G}^{-}_{\phi^2} + \mathcal{G}^{+*}_{\phi^2}\right) \right] \right\} |g\rangle \langle e| \nonumber\\
    &+& \frac{1}{2}\left\{ \e^{i\alpha} + \lambda^{2}_{\phi^2} \left[\e^{-i\alpha}\mathcal{C}^{+}_{\phi^2} - \e^{i\alpha} \left(\mathcal{G}^{+}_{\phi^2} + \mathcal{G}^{-*}_{\phi^2}\right) \right] \right\} |e\rangle \langle g|.
    \label{rho_I_int}
\end{eqnarray}
By the traceless condition on $\lambda_{\phi^2}$ we have
\begin{eqnarray}
    \mathcal{F}^{\mp}_{\phi^2} - 2\mathbf{Re}(\mathcal{G}^{\pm}_{\phi^2}) = 0.
    \label{rel_traceless_rho_I_int}
\end{eqnarray}

Now, by applying the second Hadamard to the density matrix given by Eq.~(\ref{rho_I_int}), we obtain (omitting the superscript ``out'')
\begin{eqnarray}
    \hat{\rho}_{I,\phi^2} &=& \Big\{ \cos^2{\frac{\alpha}{2}} + \frac{\lambda^2_{\phi^2}}{4} \Big[ \mathcal{F}_{\phi^2}^{-} + \mathcal{F}_{\phi^2}^{+} - 8\mathbf{Re}(\mathcal{G}_{\phi^2}^{-})\cos^2{\frac{\alpha}{2}} + 2\mathcal{C}_{\phi^2}^{-} \cos{\alpha} \Big] \Big\} |g\rangle \langle g| + \nonumber\\
    &+& \Big\{ \sin^2{\frac{\alpha}{2}} + \frac{\lambda^2_{\phi^2}}{4} \Big[ \mathcal{F}_{\phi^2}^{-} + \mathcal{F}_{\phi^2}^{+} - 8\mathbf{Re}(\mathcal{G}_{\phi^2}^{-})\sin^2{\frac{\alpha}{2}} - 2\mathcal{C}_{\phi^2}^{-} \cos{\alpha} \Big] \Big\} |e\rangle \langle e| +\nonumber\\
    &+&\Big\{ -\frac{i}{2}\sin{\alpha} + \frac{\lambda^2_{\phi^2}}{4} \Big[ \mathcal{F}_{\phi^2}^{-} - \mathcal{F}_{\phi^2}^{+} + 2i\sin{\alpha \left( 2\mathbf{Re}(\mathcal{G}_{\phi^2}^{-}) - \mathcal{C}_{\phi^2}^{-}\right)}\Big] \Big\} |e\rangle\langle g| + \nonumber\\
    &+& \Big\{ +\frac{i}{2}\sin{\alpha} + \frac{\lambda^2_{\phi^2}}{4} \Big[ \mathcal{F}_{\phi^2}^{-} - \mathcal{F}_{\phi^2}^{+} - 2i\sin{\alpha \left( 2\mathbf{Re}(\mathcal{G}_{\phi^2}^{-}) - \mathcal{C}_{\phi^2}^{-}\right)}\Big] \Big\} |g\rangle\langle e|,
    \label{rhoI final}
\end{eqnarray}
where it must obey the following relation to be traceless in $\lambda^2_{\phi^2}$,
\begin{eqnarray}
    \mathbf{Re}(\mathcal{G}_{\phi^2}^{-}) = \frac{1}{4}\left( \mathcal{F}_{\phi^2}^{-} + \mathcal{F}_{\phi^2}^{+}\right).
    \label{tracelessrelI}
\end{eqnarray}

The expression given by Eq. (\ref{rhoI final}) represents the density matrix of the quantum interferometric circuit. Through it, it is possible to calculate the probability of the internal states, the interferometric visibility and the quantum coherence; and this is what we will do in the subsequent sections.

\subsubsection{Probability of internal states}

For the case of the quantum interferometric circuit with quadratic coupling, we can also investigate the probabilities of finding the internal states of the detector, as was investigated in the setup of the accelerated qubit. In this way, the definition of the probability of finding the ground state and the excited state, respectively, is written as
\begin{eqnarray}
    P^{g}_{I,\phi^2} &=& \langle g|\hat{\rho}_{I,\phi^2}|g\rangle, \\
    P^{e}_{I,\phi^2} &=& \langle e|\hat{\rho}_{I,\phi^2}|e\rangle,
\end{eqnarray}
where these expressions provide us
\begin{eqnarray}
    P^{g}_{I,\phi^2} &=&  \cos^2{\frac{\alpha}{2}} + \frac{\lambda^2_{\phi^2}}{4} \Big[ \mathcal{F}_{\phi^2}^{-} + \mathcal{F}_{\phi^2}^{+} - 8\mathbf{Re}(\mathcal{G}_{\phi^2}^{-})\cos^2{\frac{\alpha}{2}} + 2\mathcal{C}_{\phi^2}^{-} \cos{\alpha} \Big], \\
    P^{e}_{I,\phi^2} &=& \sin^2{\frac{\alpha}{2}} + \frac{\lambda^2_{\phi^2}}{4} \Big[ \mathcal{F}_{\phi^2}^{-} + \mathcal{F}_{\phi^2}^{+} - 8\mathbf{Re}(\mathcal{G}_{\phi^2}^{-})\sin^2{\frac{\alpha}{2}} - 2\mathcal{C}_{\phi^2}^{-} \cos{\alpha} \Big],
\end{eqnarray}
By employing the relation $\mathcal{F}^{\pm}_{\phi^{2}} = \sigma\overline{\mathcal{R}}^{\pm}_{\phi^2}$, substituting Eq.~(\ref{Solution_Integral_C}), performing the corresponding algebraic manipulations, and taking the limit $\sigma \gg 1$, we obtain the following expressions:
\begin{eqnarray}
    P^{g}_{I,\phi^2} &=&  \cos^2{\frac{\alpha}{2}} - (1+\overline{a}^2)\frac{\sigma\Lambda^2}{96\pi^3} \coth{\left( \frac{\pi}{\overline{a}}\right)} \cos{\alpha}, \\
    P^{e}_{I,\phi^2} &=& \sin^2{\frac{\alpha}{2}} + (1+\overline{a}^2)\frac{\sigma\Lambda^2}{96\pi^3} \coth{\left( \frac{\pi}{\overline{a}}\right)} \cos{\alpha},
\end{eqnarray}
which are the final expressions for finding the probabilities of the internal states in the interferometric circuit with quadratic coupling. Note that $P^{\mathrm{total}}_{I,\phi^2} = P^{g}_{I,\phi^{2}} + P^{e}_{I,\phi^{2}} = 1$.

Similarly, for the linear case\footnote{See Eq.~(4.8) of ref. \cite{pedro2025mitigating} considering the massless scalar quantum field.}, the probabilities of the internal states are written as follows
\begin{eqnarray}
    P^{g}_{I,\phi} &=&  \cos^2{\frac{\alpha}{2}} - \frac{\sigma\lambda^2}{8\pi} \coth{\left( \frac{\pi}{\overline{a}}\right)} \cos{\alpha}, \\
    P^{e}_{I,\phi} &=& \sin^2{\frac{\alpha}{2}} + \frac{\sigma\lambda^2}{8\pi} \coth{\left( \frac{\pi}{\overline{a}}\right)} \cos{\alpha},
\end{eqnarray}
where we see again that the change between both linear and quadratic coupling is given only through the coupling constant and the factor $(1+\overline{a}^2)/12\pi^2$.

\subsubsection{Visibility and quantum coherence }

At this stage, we are equipped to extract information regarding the interference pattern generated by the quantum interferometric circuit. The interferometric visibility (commonly referred to as visibility) serves as a quantitative measure of the contrast in interference arising from superposition. Visibility is formally defined as the ratio between the amplitude of the interference pattern and the sum of the individual intensities. Mathematically, it is expressed as:
\begin{eqnarray}
    \mathcal{V}_{I,\phi^2} = \frac{P^{\mathrm{g, max}}_{I,\phi^2} - P^{\mathrm{g, min}}_{I,\phi^2}}{P^{\mathrm{g, max}}_{I,\phi^2} + P^{\mathrm{g, min}}_{I,\phi^2}}.
\end{eqnarray}
Here, the probability $P^{\mathrm{g}}_{I,\phi^2}$ reaches its maximum (minimum) value when $\alpha = 0$ ($\alpha = \pi$). Accordingly, by considering the limit of long interaction time, i.e., $\sigma \gg 1$, we obtain:
\begin{eqnarray}
    \mathcal{V}_{I,\phi^2} &\approx& 1 - (1+\overline{a}^2)\frac{\sigma \Lambda^2}{48\pi^3} \coth{\left(\frac{\pi}{\overline{a}}\right)},
    \label{visibility_quadratic}
\end{eqnarray}
which represents the visibility for the interferometric circuit quadratically coupled with the scalar field. And for the case of linear coupling, we obtain
\begin{eqnarray}
    \mathcal{V}_{I,\phi} &\approx& 1 - \frac{\sigma \lambda^2}{4\pi} \coth{\left(\frac{\pi}{\overline{a}}\right)}.
    \label{visibility_linear}
\end{eqnarray}

Additionally, we can calculate the $l^1$ norm coherence by summing the off-diagonal elements of the reduced density matrix given by Eq.~(\ref{rhoI final}), thus we obtain
\begin{eqnarray}
    \mathcal{Q}^{l^1}_{I,\phi^2} &\approx& \sin{(\alpha)}\left[1 - (1+\overline{a}^2)\frac{\sigma \Lambda^2}{48\pi^3} \coth{\left(\frac{\pi}{\overline{a}}\right)}\right],
    \label{coerence_Interf_quadratic}
\end{eqnarray}
for the case of quadratic coupling and 
\begin{eqnarray}
    \mathcal{Q}^{l^1}_{I,\phi} &\approx& \sin{(\alpha)}\left[1 - \frac{\sigma \lambda^2}{4\pi} \coth{\left(\frac{\pi}{\overline{a}}\right)}\right],
    \label{coerence_Interf_linear}
\end{eqnarray}
for the case of linear coupling. Once again, we observe that the distinction between linear and quadratic coupling arises solely through the coupling constant and the factor $(1 + \overline{a}^2)/12\pi^2$.

\subsubsection{Which--path distinguishability circuit}

In this section, we will adopt a modified setup, intending to extract which-path information. To achieve this, the second Hadamard gate is removed from the quantum interferometric setup, and two detectors are introduced. In this scenario, the inclusion of the two path detectors enables the quadratic interaction between the qubit and a massless scalar field to reveal the path taken by the qubit.

Specifically, the internal state of the qubit is prepared as $|g\rangle$ for qubits detected along path A by detector A. Conversely, qubits not detected by A are inferred to have traveled along path B. In this manner, the which-path information is effectively encoded in the internal states of the qubit. In this configuration, the reduced density matrix describing the system after the detector–field interaction to is given by Eq.~(\ref{rho_I_int}).

A key quantity in this system is the which--path distinguishability, which characterizes the particle--like behavior of the qubit. This quantity is quantified by the following expression:
\begin{equation}
 \mathcal{D}_{\phi^2} = \frac{|w_{A,\phi^2} - w_{B,\phi^2}|}{w_{A,\phi^2} + w_{B,\phi^2}},
 \label{Dis1}
\end{equation}
where $w_{A,\phi^2}$ denotes the probability of detecting the qubit with detector A (corresponding to path A), and $w_{B,\phi^2}$ represents the probability of detection by detector B (corresponding to path B). By employing Eqs.~(\ref{rho_I_int}) and (\ref{rel_traceless_rho_I_int}), the expression can be written as:
\begin{eqnarray}
 w_{A,\phi^2} &=& \frac{1}{2} + \frac{\lambda^2_{\phi^2}}{2} \left(\mathcal{F}_{\phi^2}^- - \mathcal{F}_{\phi^2}^+\right), \\
 w_{B,\phi^2} &=& \frac{1}{2} + \frac{\lambda^2_{\phi^2}}{2} \left(\mathcal{F}_{\phi^2}^+ - \mathcal{F}_{\phi^2}^-\right).
\end{eqnarray}

By using $\mathcal{F}_{\phi^2}^\pm = \sigma \overline{\mathcal{R}}_{\phi^2}^\pm$ and Eq. \eqref{Dis1}, the which--path distinguishability is
\begin{equation}
 \mathcal{D}_{\phi^2} = (1+\overline{a}^2)\frac{\sigma\Lambda^2}{24\pi^3} + \mathcal{O}(\Lambda^{4}).
 \label{DisFinal_quadratic}
\end{equation}
and for the linear case\footnote{See Eq.~(4.21) of ref. \cite{pedro2025mitigating}, and apply the limit for the massless field to obtain which-path distinguishability for the linear case.}, we obtain
\begin{equation}
 \mathcal{D}_{\phi} = \frac{\sigma\lambda^2}{2\pi} + \mathcal{O}(\lambda^{4}).
 \label{DisFinal_linear}
\end{equation}
Once again, it becomes evident that the distinction between linear and quadratic coupling manifests exclusively through the coupling constant and the multiplicative factor $(1 + \overline{a}^2)/12\pi^2$. Therefore, we can see from these results that for quadratic coupling [Eq.~(\ref{DisFinal_quadratic})], which-path distinguishability depends on the acceleration, unlike the case of linear coupling [Eq.~(\ref{DisFinal_linear})].

\subsubsection{Complementarity relation}

The wave–particle relation quantifies the fundamental trade-off between these two quantities. Specifically, any attempt to acquire which-path information (i.e., increase in distinguishability $\mathcal{D}_{\phi^{2}}$) inevitably results in a loss of coherence in the interference pattern (i.e., decrease in visibility $\mathcal{V}_{I,\phi^{2}}$), and vice versa. This trade-off is rigorously expressed by the inequality established in the seminal works of Englert and Zeilinger \cite{Englert1996,Englert1999,Zeilinger1999Experiment}, given by
\begin{equation} 
    C_{I,\phi^{2}} = \mathcal{V}^2_{I,\phi^{2}} + \mathcal{D}^2_{\phi^{2}} \leq 1.
\label{Complementarity1_quadratic}
\end{equation}
Eq.~\eqref{Complementarity1_quadratic} represents the duality relation specific to a Ramsey interferometer \cite{Ramsey1950}. The equality in this relation holds strictly under the condition that both the qubit and the quantum field are described by pure states. In all other cases, where either or both subsystems are in mixed states, the inequality becomes strict. Importantly, Eq.~\eqref{Complementarity1_quadratic} encapsulates the fundamental principle of complementarity by establishing a mutually exclusive relationship between the wave-like and particle-like behaviors exhibited by the qubit.

Finally, substituting Eqs. (\ref{visibility_quadratic}) and (\ref{DisFinal_quadratic}) into (\ref{Complementarity1_quadratic}), we concluded
\begin{eqnarray}
    C_{I,\phi^2} &\approx& 1 - (1+\overline{a}^2)\frac{\sigma \Lambda^2}{24\pi^3} \coth{\left(\frac{\pi}{\overline{a}}\right)} + \mathcal{O}(\Lambda^4),
    \label{ComplementarityFinal_quadratic}
\end{eqnarray}
where we take the limit for a long time of interaction, i.e. $\sigma \gg 1$. And similarly, for the case of linear coupling\footnote{The complementarity relation for the linear coupling case is obtained by taking the limit for the massless field in Eq.~(4.23) of ref. \cite{pedro2025mitigating}.}, we get
\begin{eqnarray}
    C_{I,\phi} &\approx& 1 - \frac{\sigma \lambda^2}{2\pi} \coth{\left(\frac{\pi}{\overline{a}}\right)} + \mathcal{O}(\lambda^4).
    \label{ComplementarityFinal_linear}
\end{eqnarray}
Once again, it becomes evident that the difference between linear and quadratic coupling is manifested through the coupling constant and the multiplicative factor~$(1 + \overline{a}^2)/12\pi^2$. This highlights that the qualitative behavior of the system remains structurally similar across both coupling regimes, with quantitative deviations arising solely from these specific parameters.

\subsubsection{Numeric results}

\begin{figure}
    \centering
    \includegraphics[width=0.3\linewidth]{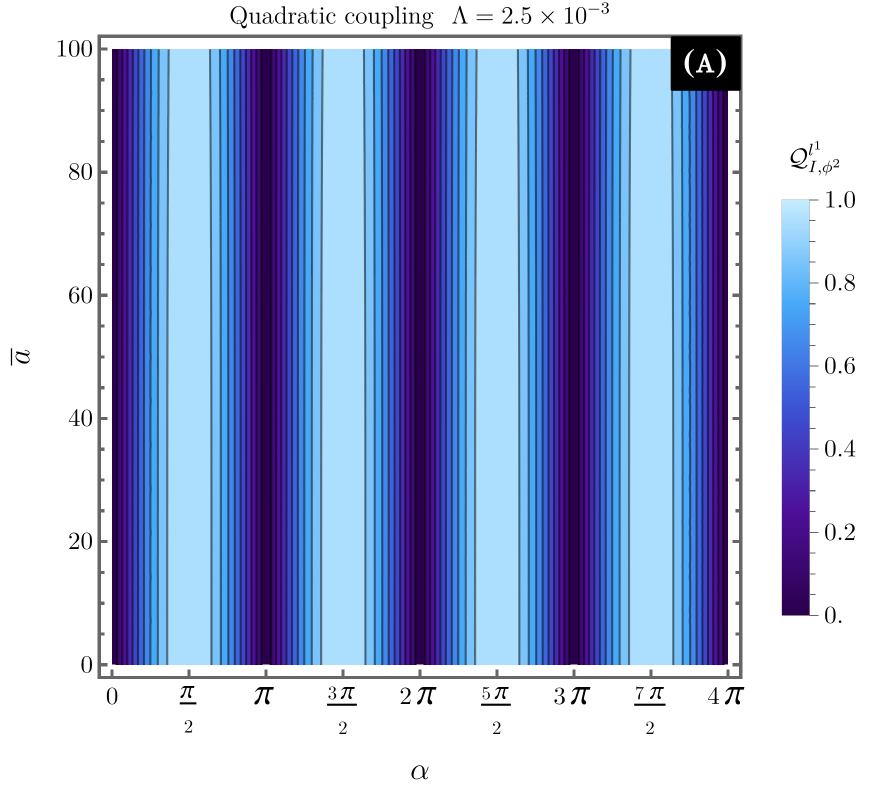}
    \includegraphics[width=0.3\linewidth]{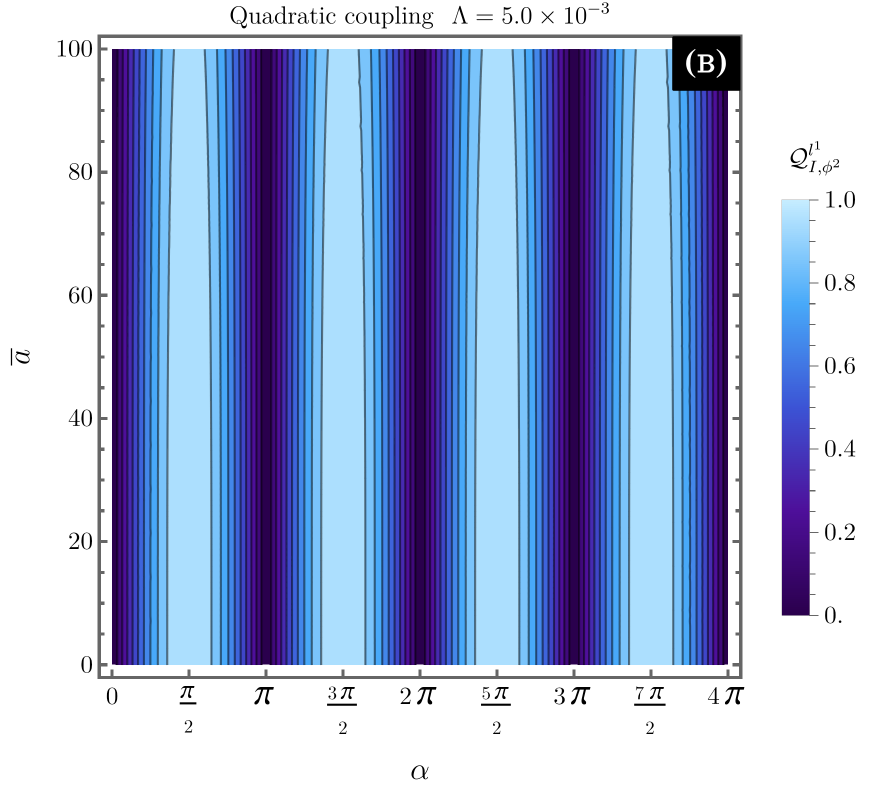}
    \includegraphics[width=0.3\linewidth]{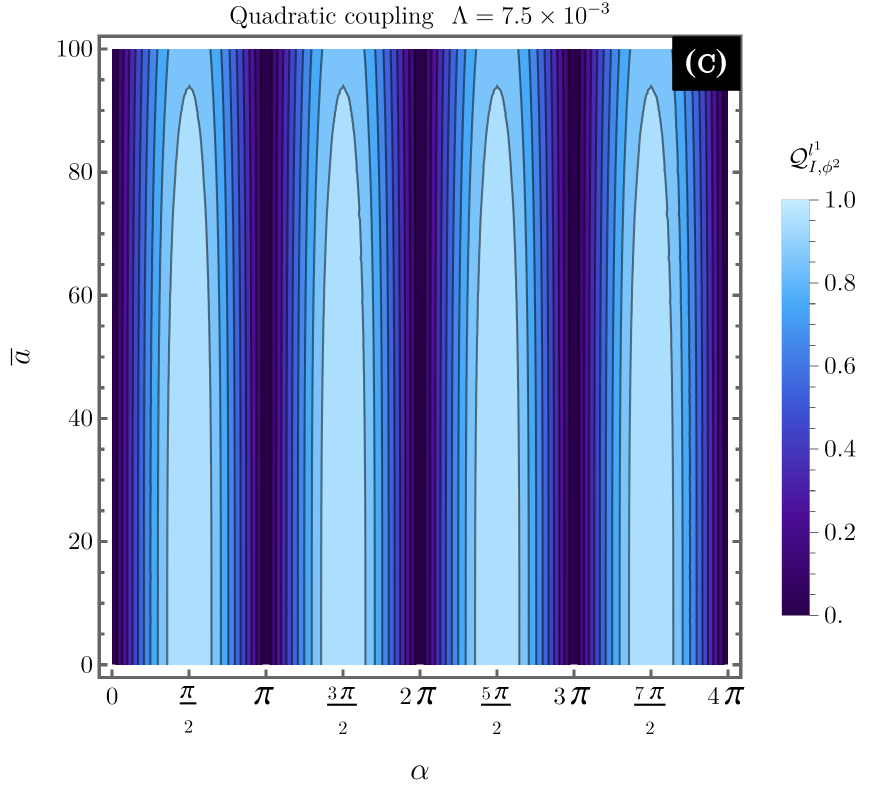}
    \includegraphics[width=0.3\linewidth]{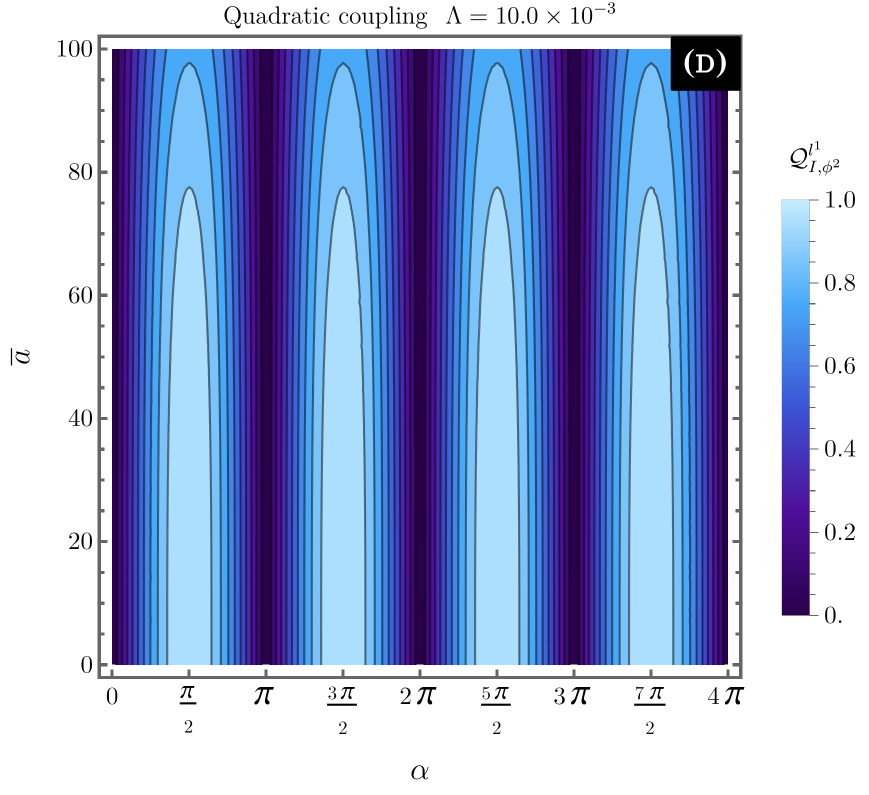}
    \includegraphics[width=0.3\linewidth]{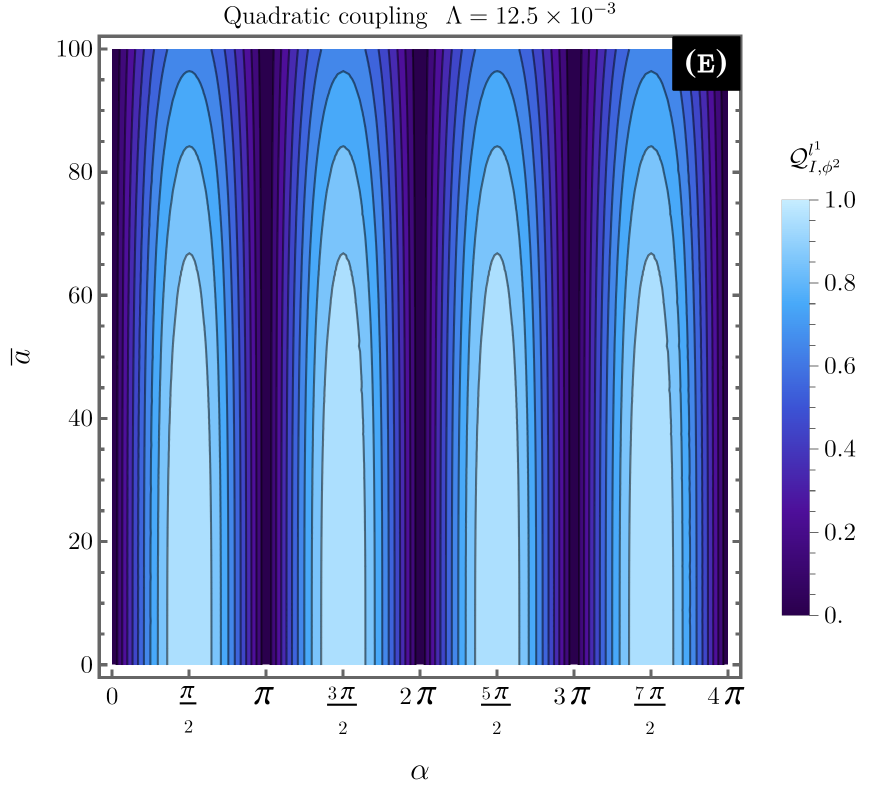}
    \includegraphics[width=0.3\linewidth]{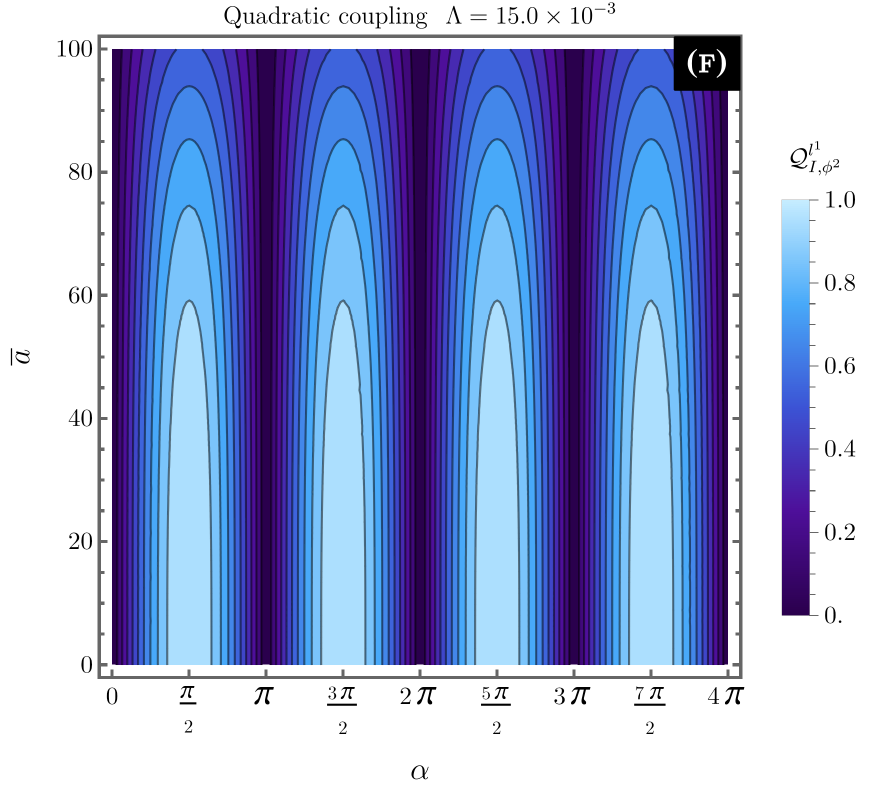}
    \caption{Quantum coherence $\mathcal{Q}^{l^1}_{I,\phi^2}$ (for quadratic coupling) as a function of $\overline{a}$ and $\alpha$, for several values of $\Lambda$. We adopted $\sigma = 10$.}
    \label{Fig4}
\end{figure}

The numerical results depicted in Fig.~\ref{Fig4}(a)–~\ref{Fig4}(f) correspond to the scenario of quadratic coupling. For high acceleration values and phase parameters set to half-integer multiples of $\pi$, the results clearly indicate a degradation of quantum coherence. Moreover, under these specific conditions, it is observed that increasing the value of $\Lambda$ leads to broader and less well-defined interference fringes in the coherence profiles of the quantum interferometric circuit, signaling a reduction in the system’s coherence.

\begin{figure}
    \centering
    \includegraphics[width=0.3\linewidth]{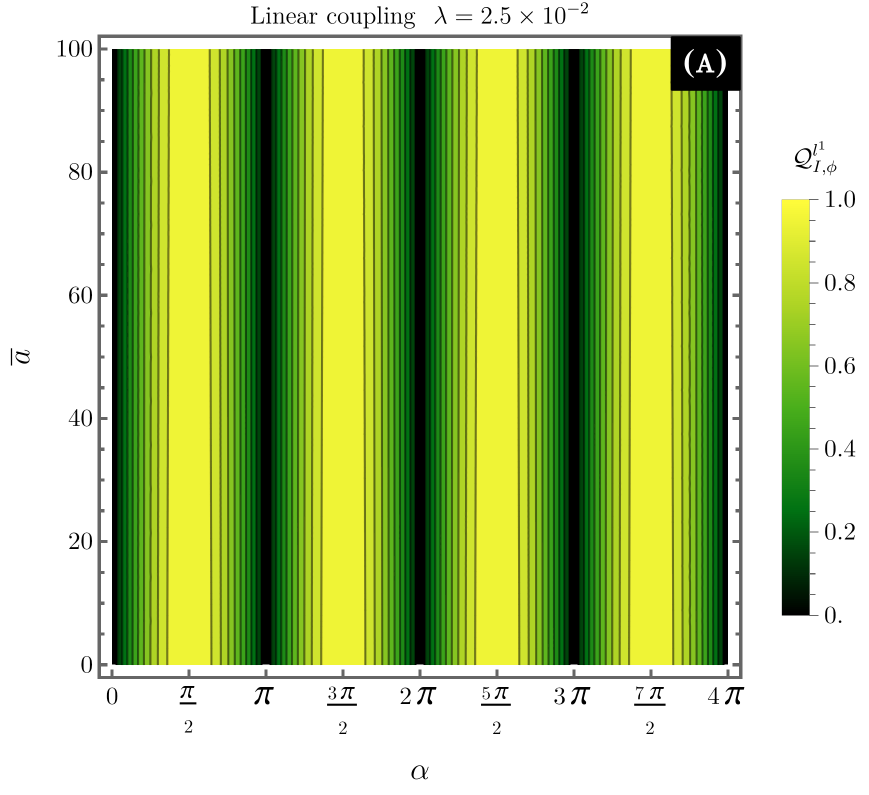}
    \includegraphics[width=0.3\linewidth]{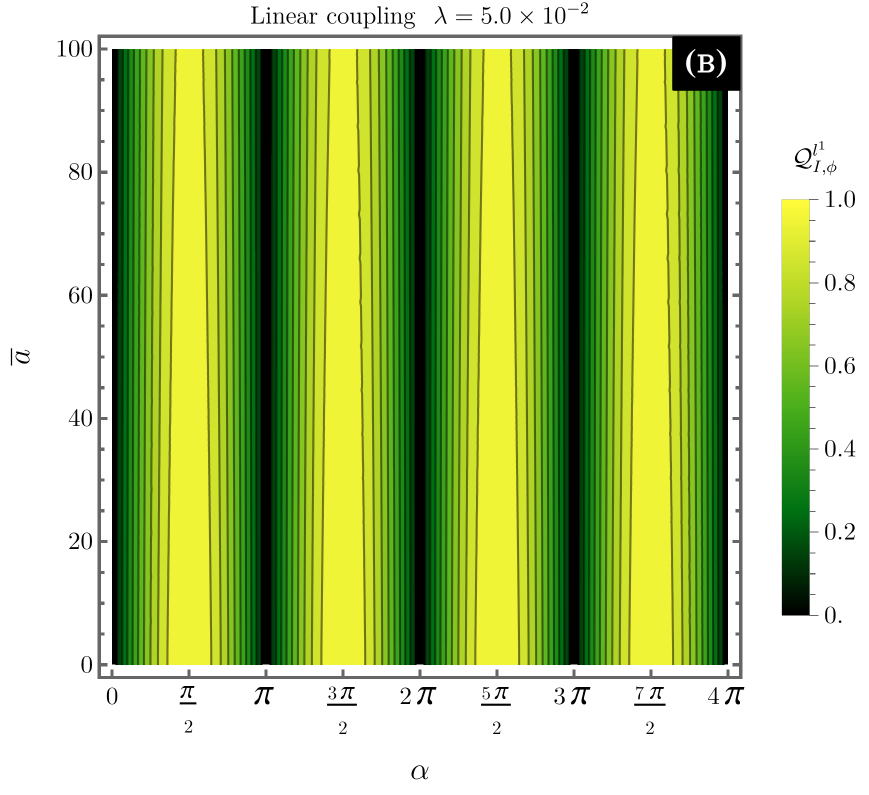}
    \includegraphics[width=0.3\linewidth]{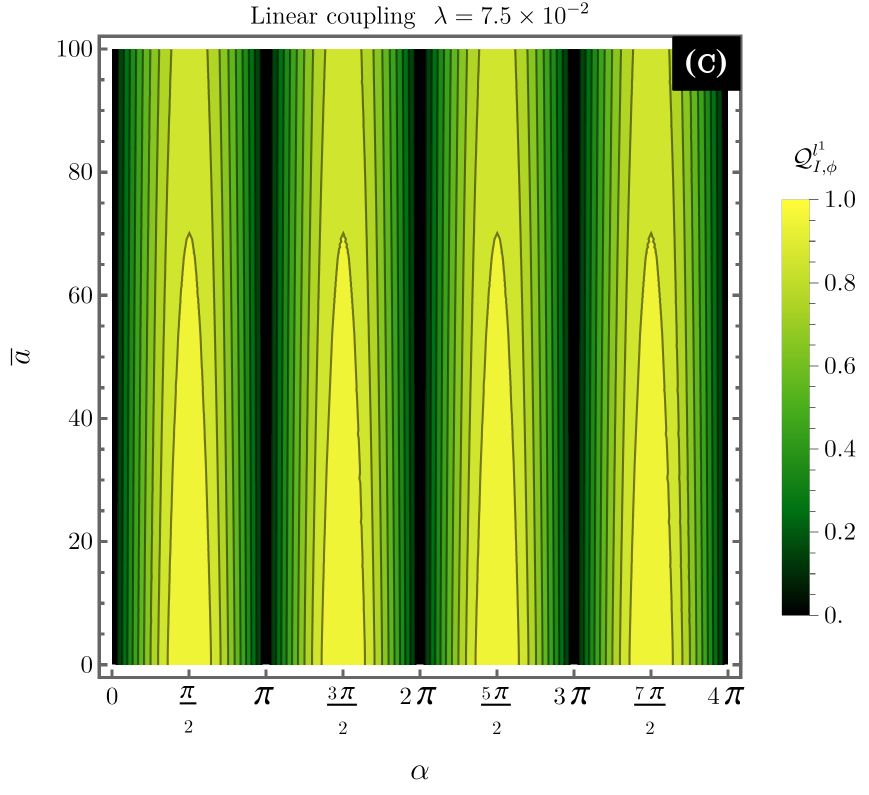}
    \includegraphics[width=0.3\linewidth]{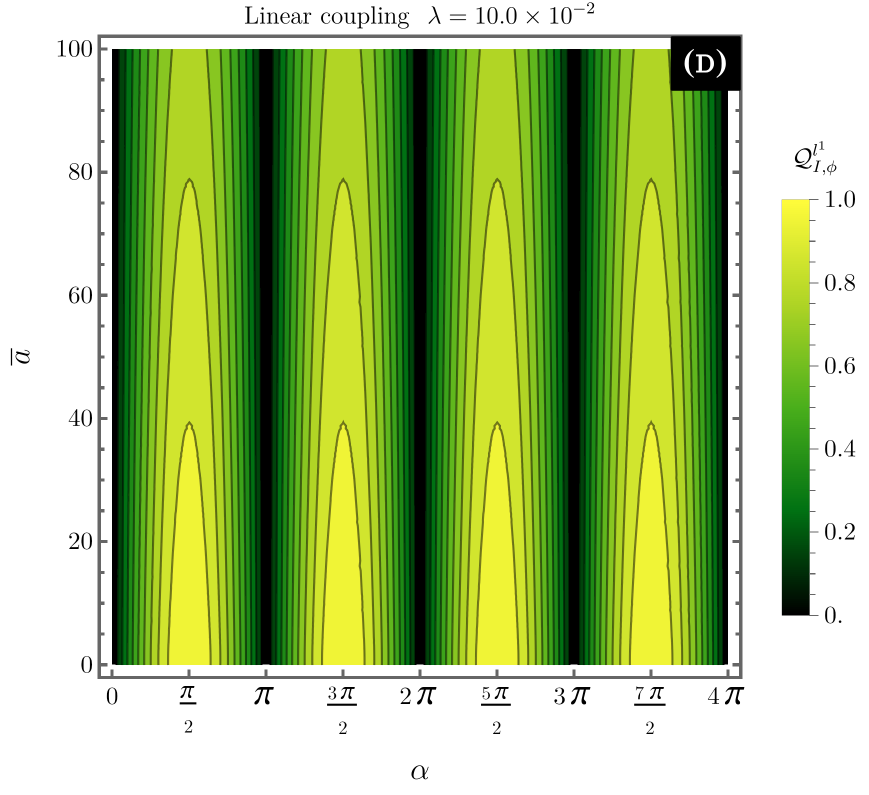}
    \includegraphics[width=0.3\linewidth]{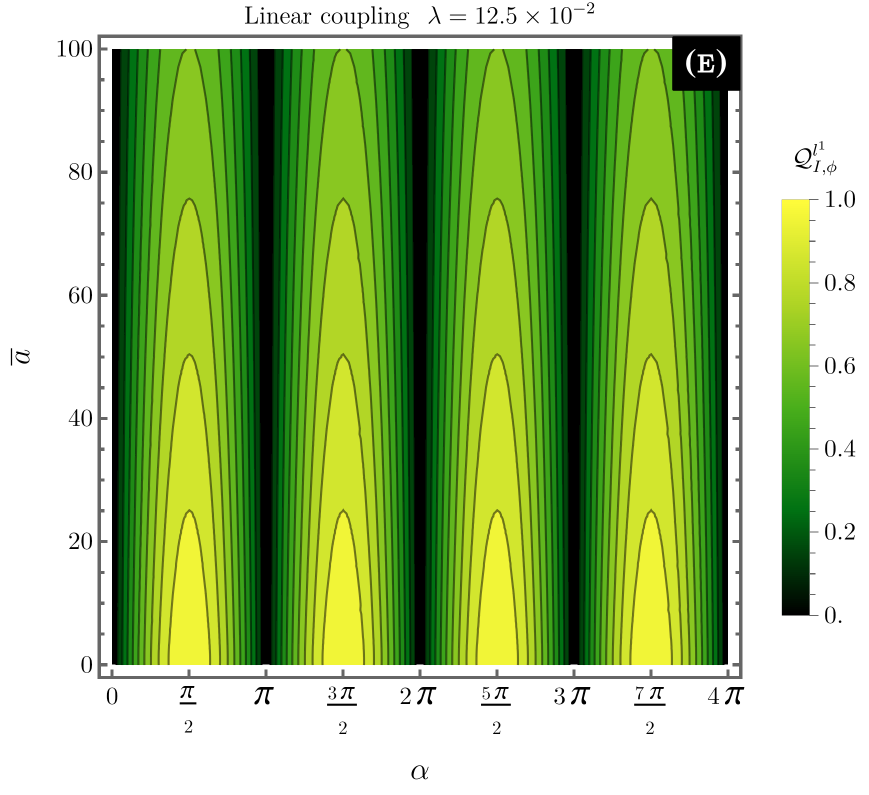}
    \includegraphics[width=0.3\linewidth]{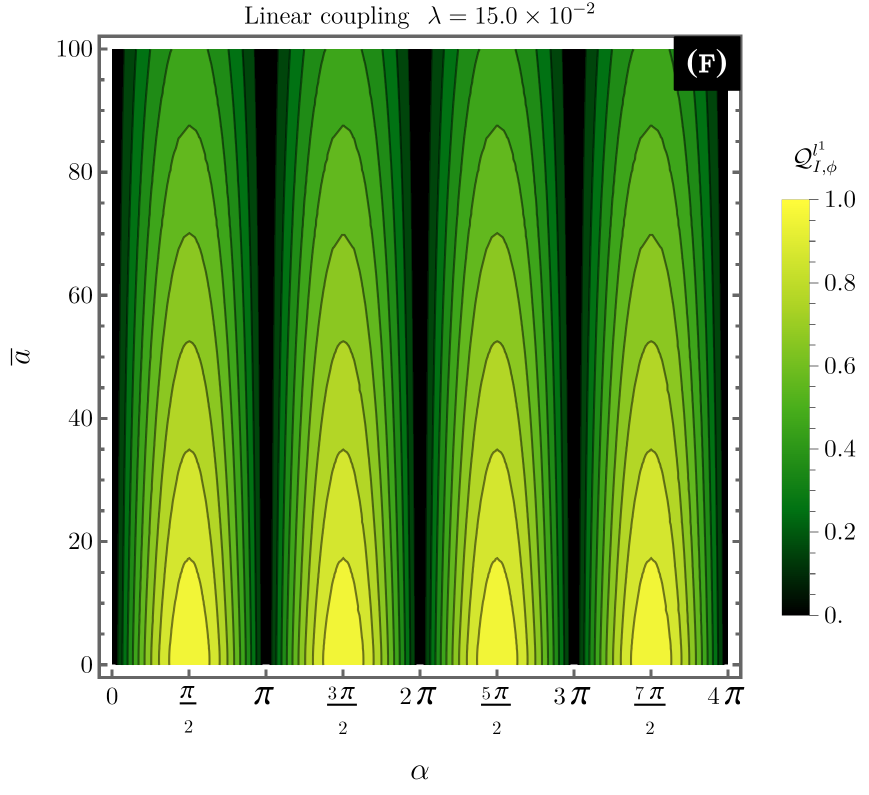}
    \caption{Quantum coherence $\mathcal{Q}^{l^1}_{I,\phi}$ (for linear coupling) as a function of $\overline{a}$ and $\alpha$, for several values of $\lambda$. Note that in all plots, we assume $\lambda$ to be 10 times larger than $\Lambda$ to visualize significant effects. We adopted $\sigma = 10$.}
    \label{Fig5}
\end{figure}

Similarly, for the linear coupling case we have Fig. Fig.~\ref{Fig5}(a)–~\ref{Fig5}(f), we also note that for larger values of $\lambda$, the quantum coherence profiles are less sharp fringes. On the other hand, it is important to emphasize that for the linear case we applied for $\lambda$ values 10 times larger than $\Lambda$ in order to visualize the significant effects for the linear case. This makes the difference in strength between the couplings in question clear.

\begin{figure}
    \centering
    \includegraphics[width=0.475\linewidth]{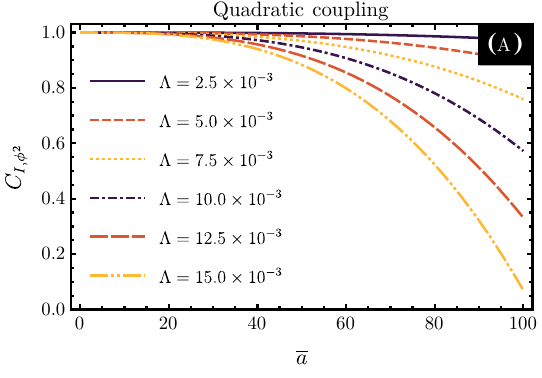}
    \includegraphics[width=0.49\linewidth]{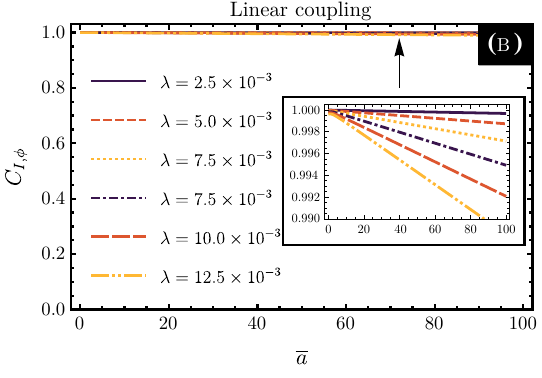}
    \includegraphics[width=0.475\linewidth]{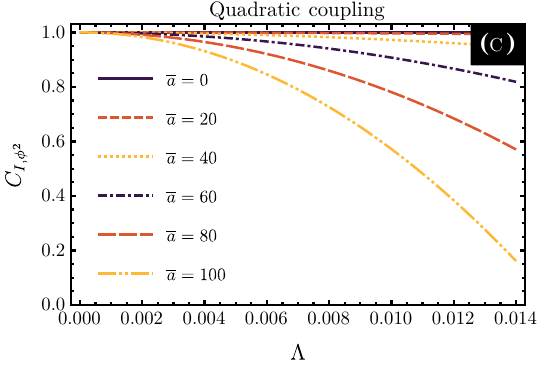}
    \includegraphics[width=0.49\linewidth]{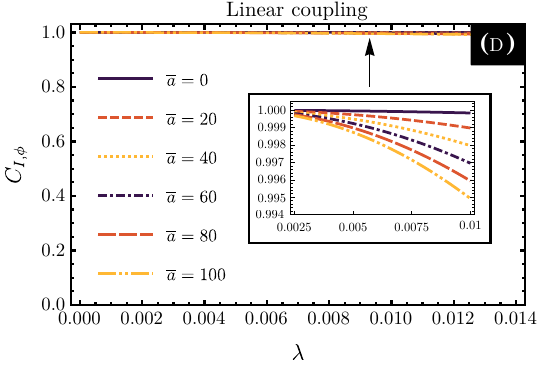}
    \includegraphics[width=0.49\linewidth]{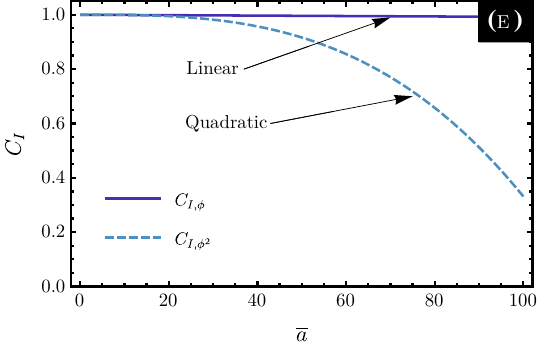}
    \caption{The representation of $C_{I}$ as a function of the parameter $\overline{a}$: \textbf{(a)} for quadratic coupling with different values of $\Lambda$, and \textbf{(b)} for linear coupling with different values of $\lambda$. Dependence of $C_{I}$ on \textbf{(c)} $\Lambda$ and \textbf{(d)} $\lambda$ for different values of $\overline{a}$. \textbf{(e)} Plot of $C_{I,\phi^2}$ and $C_{I,\phi}$ as functions of $\overline{a}$ with the implicit assumption that $\Omega = 1$. In all cases, we adopted  $\sigma = 10$.}
    \label{Fig6}
\end{figure}

In Fig. \ref{Fig6}, we numerically analyze the effects of quadratic coupling on the wave--particle duality under the Unruh effect. This is done by means of the complementarity relation, which makes use of which--path distinguishability and interferometric visibility. In this way, in Fig.~\ref{Fig6}(a) and Fig.~\ref{Fig6}(b), we plot the complementarity relation for both couplings as a function of the acceleration for different values of their respective coupling constants. Through these plots, it can be noted that for quadratic coupling the loss of wave--particle information occurs much faster when compared to the linear case.

However, the same effect is observed when we compare the complementarity relation as a function of the coupling constant $\Lambda$ [Fig.~\ref{Fig6}(c)], and as a function of the coupling constant $\lambda$ [Fig.~\ref{Fig6}(d)], both considering different values of the parameter $\overline{a}$. That is, it is verified that considering a quadratic coupling the information degradation of the wave-particle duality is amplified when compared to the linear coupling.

Thus, to further clarify this discussion, observe Fig.~\ref{Fig6}(e), where we represent in the same plot the complementarity relation as a function of $\overline{a}$ for both couplings in question, where by implicitly assuming that $\Omega = 1$ we can assume that $\Lambda = \lambda = 1.25 \times 10^{-2}$. Now, it is possible to easily visualize the difference between the degradation of wave-particle duality information when the system reaches high accelerations.
Revealing that when considering a detector at high accelerations quadratically coupled with a massless scalar field, the loss of information happens more quickly compared to the case when the detector is coupled linearly with a field of the same type.

\section{Summary and conclusion\label{sec:4}}

We investigate the information degradation of quantum systems under the Unruh effect through detectors quadratically coupled with a massless  scalar field, and compare the findings with the case of linear coupling. The results show that, under high acceleration, coherence and probability amplitudes of the internal states decay more rapidly in the case of quadratic coupling. This stronger degradation is attributed to the coupling constant and the additional factor $(1 + \overline{a}^2)/12\pi^2$ that arises in the quadratic interaction.

More specifically, a detector quadratically coupled with a scalar field responds more due to the larger dimensionality of its coupling constant, which at $d = 4$ we have $\Delta_{\phi^2} = 1$, contrasting with $\Delta_{\phi} = 0$ for scalar fields~\cite{gray2018scalar} (for more details see the second paragraph of Sect. \ref{Sec IIA}). Consequently, the change in the interaction structure implies the emergence of the factor $(1 + \overline{a}^2)/12\pi^2$, as shown in Eq.~(\ref{relation_linear_quadratic}), this factor is responsible for the difference in response between the linear and quadratic cases.

For the setup of the interferometric circuit, our findings showed that when investigating coherence as a function of phase we obtain that the bright interference fringes degrade as we increase the coupling constant. Furthermore, this fringe degradation effect occurs more strongly for quadratic coupling. For linear coupling one should use a coupling constant 10 times larger to obtain a level of fringe degradation similar to the quadratic case.

By employing interferometric visibility and which--path distinguishability, we established the complementarity relation, which enabled us to investigate the influence of quadratic coupling on the manifestation of wave--particle duality in the presence of the Unruh effect. The results revealed an amplification degradation of both wave--like and particle--like information when quadratic coupling is considered. This suggests that the nature of the coupling between the scalar field and the detector plays a significant role in amplifying the effective thermal environment induced by high acceleration.

Quadratic coupling amplifies the Unruh effect by modifying the interaction structure, allowing the simultaneous absorption of multiple quanta. Unlike linear coupling, which involves direct interactions with individual field excitations, quadratic coupling involves higher--order terms that increase the spectral density and introduce new excitation channels, resulting in a stronger response function and higher transition rates.
 
In more detail, for the detector to be excited, the sum of the frequencies of two quanta must equal the detector's energy gap \cite{unruh1976}. This higher-order interaction opens new excitation channels not available in the linear case, resulting in higher transition rates. Mathematically, the Wightman function for quadratic coupling is proportional to the square of the Wightman function for linear coupling ($\mathcal{W}_{\phi^2} \propto \mathcal{W}_\phi^2$), reflecting this amplification of correlations. In fermionic fields, quadratic coupling is even necessary to preserve the conservation of fermionic number, allowing the creation or annihilation of particle-antiparticle pairs \cite{MartinezRenormalized2016}.

Therefore, these results demonstrate that the nature of the coupling between the detector and the field plays a fundamental role in determining the degree of information preservation in accelerated quantum systems. Thus, our findings advance the fundamental understanding of RQI. Overall, these results deepen our understanding of quantum information preservation and open doors to promising technological prospects.

\section*{Acknowledgments}

P.~H.~M.~B acknowledges the Brazilian funding agency CAPES for financial support through grant No.~88887.674765/2022-00 (Doctoral Fellowship – CAPES). P.~R.~S.~C would like to thank the Brazilian funding agencies CAPES and CNPq (Grant: Produtividade 306130/2022-0) for financial support. The authors thank the Referee for helpful comments.

\section*{Conflicts of interest/competing interest}

The authors declared that there is no conflict of interest in this manuscript.

\section*{Data Availability Statement}

This article has no associated data or the data will not
be deposited.

\section*{Code Availability Statement}

This article has no associated code or the code will not
be deposited.

\appendix

\section{Brief review of scalar field theory}

In this appendix, we will briefly review quantum scalar field theory in order to clarify and fill any remaining gaps regarding the use of field theory in the main body of this work (for more details see references \cite{lancaster2014quantum, birrell1984quantum}). In this way, first consider a real, massless, scalar field $\phi(x)$, it is well known that the corresponding Lagrangian density is given by \cite{lancaster2014quantum}
\begin{eqnarray}
    \mathcal{L}(x) = \frac{1}{2} \partial^{\mu}\phi(x) \partial_{\mu}\phi(x),  
\end{eqnarray}
and furthermore, the canonical conjugate momentum corresponding to $\mathcal{L}(x)$ is written as
\begin{eqnarray}
    \pi(x) = \frac{\partial\mathcal{L}(x)}{\partial[\partial_{0}\phi(x)]}  = \partial_{0}\phi(x).
\end{eqnarray}
Thus, using the Legendre transform, we can construct the Hamiltonian, i.e.
\begin{eqnarray}
    \mathcal{H}_{\phi} = \int \d^{3}x \frac{1}{2} [(\partial_{t}\phi)^2 + (\nabla\phi)^2],
\end{eqnarray}
where the equations of motion lead us to the Klein--Gordon equation, and therefore we have
\begin{eqnarray}
    \Box\phi(x) = (\partial_{t}^2 + \nabla^2)\phi(x) = 0,
\end{eqnarray}
where this equation has well-known solutions, which are plane wave solutions, and thus we obtain
\begin{eqnarray}
    u_{\mathbf{k}}(x) = \frac{1}{\sqrt{2}} \e^{-ikx},
\end{eqnarray}
with $kx = k_{\mu} x^{\mu} = k_{0} x^{0} - \mathbf{k} \cdot \mathbf{x}$ and $k^{0} = \vert \mathbf{k} \vert$. These are called positive frequency modes with respect to $t$. One can then define the inner product,
\begin{eqnarray}
    \left( \phi_{1},\phi_{2}\right) = -i \int \d^{3}x \Big( \phi_{1}(x) \partial_{t} \phi_{2}^{*}(x) - \partial_{t}[\phi_{1}(x)]\phi_{2}^{*}(x)\Big).
\end{eqnarray}

Thus, it can be seen that the positive frequency modes of the equation form a complete orthogonal basis and therefore we can expand the field, and thus we have
\begin{eqnarray}
    \phi(t,\mathbf{x}) = \int \frac{\d^{3}x}{(2\pi)^3 2k^{0}} \Big( a_{\mathbf{k}}u_{k}(x) + a_{\mathbf{k}}^{\dagger}u_{k}^{*}(x)\Big),
    \label{expansion_mode}
\end{eqnarray}
where $a_{\mathbf{k}}$ and $a_{\mathbf{k}}^{\dagger}$ are the annihilation and creation operators, respectively. Now, using the canonical quantization method, and impose the equal time commutation relations, we obtain
\begin{eqnarray}
 [\phi(t,\mathbf{x}),\phi(t,\mathbf{x'})] = 0, \qquad [\pi(t,\mathbf{x}),\pi(t,\mathbf{x}')] = 0, \qquad [\phi(t,\mathbf{x}),\pi(t,\mathbf{x}')] = i\delta^{3}(\mathbf{x}-\mathbf{x}').
 \label{commutation_relations}
\end{eqnarray}
Therefore, through the vacuum state $\vert 0 \rangle$ the Fock-Space representation of Hilbert space is given by
\begin{eqnarray}
    a_{\mathbf{k}} \vert 0 \rangle = 0 \quad \text{and} \quad \langle 0 \vert a_{\mathbf{k}}^{\dagger} = 0, \quad \text{for all} \quad \mathbf{k},
    \label{properties_operators}
\end{eqnarray}
where this shows that vacuum is the state that contains no particles. On the other hand, the operator $a_{\mathbf{k}}^{\dagger}$ creates a particle in state $\mathbf{k}$, $\vert 1_{\mathbf{k}}\rangle = a_{\mathbf{k}}^{\dagger} \vert 0 \rangle$, and in this way, we can construct several particle states with successive applications of the creation operator.

\section{Linear and quadratic Wightman functions}

In this appendix, we derive the vacuum correlation functions. Specifically, we demonstrate that, in the case of massless fields, all such correlators can be expressed solely in terms of the scalar Wightman function. For the linear case we have the Wightman function is written as
\begin{eqnarray}
    \mathcal{W}(x,y) = \langle 0\vert \phi(x) \phi(y) \vert 0\rangle,
\end{eqnarray}
and we can continue using the mode expansion given by Eq. (\ref{expansion_mode}), in addition we also use the commutation relations [Eq. (\ref{commutation_relations})] and the properties of the creation and annihilation operators [Eq. \ref{properties_operators}], this results in the expression
\begin{eqnarray}
    \mathcal{W}(x,y) = \int \frac{\d^4 k}{(2\pi)^3}\Theta(k^0) \delta(k^2) \e^{-ik(x-y)-k^0 \epsilon},
\end{eqnarray}
with $\Theta(\cdots)$ is the Heaviside function. Note that we explicitly include the regulator $\epsilon > 0$ and the Wightman function should be understood as a distribution in the limit $\epsilon \to 0$. Now, we define $\Delta = \vert (x-y)\vert$ and using the Lorentz transformations, we obtain 
\begin{eqnarray}
    \mathcal{W}(x,y) = \frac{1}{2(2\pi)^3} \int \d\Omega_3 \int_{0}^{\infty} k^{3} \e^{-kz} \d k,
\end{eqnarray}
where $z \equiv \epsilon + i \Delta\text{sgn}(x^{0} - y^{0}) $ and $\Omega_{3}$ is the three-dimensional angular part. Performing the last integrations over the angular part and $k$ gives
\begin{eqnarray}
    \mathcal{W}(x,y) = \frac{1}{4\pi^2}\frac{1}{[z(x,y)]^2}.
\end{eqnarray}

For quadratic coupling, according to Eq.~(\ref{defWightman}) the Wightman function for a detector interacts quadratically with the scalar field \cite{MartinezRenormalized2016, Sachs2017entanglement}, and so we write
\begin{eqnarray}
    \mathcal{W}_{\phi^2}(x,x') = \langle 0\vert :\phi^2(x)::\phi^2(y): \vert 0\rangle,
\end{eqnarray}
and making the following definitions $\phi(i) = \phi_{i}$ with $i = x, y$, we can write as
\begin{eqnarray}
    \mathcal{W}_{\phi^2}(x,x') = \langle 0\vert \phi^{+}_{x}\phi^{+}_{x}\phi^{-}_{y}\phi^{-}_{y} \vert 0\rangle = 2[\phi^{+}_{x},\phi^{-}_{y}]^2,
\end{eqnarray}
and the field is written as
\begin{eqnarray}
    \phi^{+}(x) &=& \int \frac{\d^{3}x}{(2\pi)^3 2k^{0}} a_{\mathbf{k}}u_{\mathbf{k}}(x), \\
    \phi^{-}(x) &=& \int \frac{\d^{3}x}{(2\pi)^3 2k^{0}} a^{\dagger}_{\mathbf{k}}u^{*}_{\mathbf{k}}(x),
\end{eqnarray}
in terms of positive and negative frequency components $\phi = \phi^{+} + \phi^{-}$. Now, using the normal ordering technique, we can write the following
\begin{eqnarray}
    :\phi^{2}(x): &=& \phi^2(x) - \langle 0\vert \phi^{2}(x)\vert 0\rangle, \\
    &=& (\phi^{+}_{x} + \phi^{-}_{x})^2 - [\phi^{+}_{x},\phi^{-}_{x}], \\
    &=& \phi^{+}_{x}\phi^{+}_{x} + 2\phi^{-}_{x}\phi^{+}_{x} + \phi^{-}_{x}\phi^{-}_{x},
\end{eqnarray}
and in the same way we get
\begin{eqnarray}
    :\phi^{2}(y): = \phi^{+}_{y}\phi^{+}_{y} + 2\phi^{-}_{y}\phi^{+}_{y} + \phi^{-}_{y}\phi^{-}_{y},
\end{eqnarray}
and finally we get,
\begin{eqnarray}
    \mathcal{W}_{\phi^2}(x,y) = 2[\mathcal{W}(x,y)]^2.
    \label{defWightman2}
\end{eqnarray}
This result shows, in particular, that the Wightman function for quadratic coupling depends, in a very simple way, only on the Wightman function for the case of linear coupling. For more details on the calculations, as well as for other types of couplings, see the refs.~\cite{Takagi1986vacuum, MartinezRenormalized2016, Sachs2017entanglement, gray2018scalar}.

\section{Integral calculation \texorpdfstring{$\mathcal{C}^\pm_{\phi^2}$}{C}\label{AppendixIntegralC}}

In this work, it is observed that some integrals have emerged that have quite difficult solutions. Motivated by this, we will derive here in detail the integral written as follows
\begin{eqnarray}\label{intC}
    \mathcal{C}^\pm_{\phi^2} = \int^{+\infty}_{-\infty}\d\tau \int^{+\infty}_{-\infty}\d\tau' \chi(\tau) \chi(\tau') \e^{\pm i\Omega(\tau+\tau')} \mathcal{W}^+_{\phi^2}(\tau,\tau'),
\end{eqnarray}
where $\mathcal{W}^+_{\phi^2}(\Delta\tau)$ is the Wightman function. Substituing this expression, we obtain
\begin{eqnarray}
    \mathcal{C}^\pm_{\phi^2} &=& \int^{+\infty}_{-\infty}\d\tau \int^{+\infty}_{-\infty}\d\tau' \chi(\tau) \chi(\tau') \e^{\pm i\Omega(\tau+\tau')} \Bigg\{ \frac{-1}{8\pi^4} \Bigg[\sum^{\infty}_{k=-\infty} \left( \tau - \tau' - 2i\epsilon - 2\pi ik/a\right)^{-4} + \nonumber\\
    &+& \frac{a^2}{6} \sum^{\infty}_{k=-\infty} \left( \tau - \tau' - 2i\epsilon - 2\pi ik/a\right)^{-2} \Bigg]\Bigg\},
\end{eqnarray}
and making the substitutions $s = \Delta\tau = \tau -\tau'$ and $u = \tau + \tau$, and separating the integrations, we get
\begin{eqnarray}
    \mathcal{C}^\pm_{\phi^2} &=& \frac{1}{16\pi^4}\sum^{\infty}_{k=-\infty} \int^{+\infty}_{-\infty} \exp{(-i\Omega u)}\exp{\left(\frac{-u^2}{4T^2}\right)} \d u \int^{+\infty}_{-\infty} \exp{\left(\frac{-s^2}{4T^2}\right)} \d s \nonumber\\
    &\times& \left[ \left( s - 2i\epsilon - 2\pi ik/a\right)^{-4}
    + \frac{a^2}{6} \left( s - 2i\epsilon - 2\pi ik/a\right)^{-2} \right].
\end{eqnarray}
This expression can be simplified by completing the square of the exponential of $u$, that is,
\begin{eqnarray}
    \exp{\left(-i\Omega u\right)} \exp{\left(\frac{-u}{4T^2}\right)} \quad \rightarrow \quad \exp{\left[-\left(\frac{u}{2T} + iT\Omega \right)^2\right]} \exp{(-T^2\Omega^2)},
\end{eqnarray}
and then with this we obtain the following expression
\begin{eqnarray}
    \mathcal{C}^\pm_{\phi^2} &=& \frac{1}{16\pi^4}\sum^{\infty}_{k=-\infty} \int^{+\infty}_{-\infty} \exp{\left[-\left(\frac{u}{2T} + iT\Omega \right)^2\right]} \exp{(-T^2\Omega^2)} \d u \int^{+\infty}_{-\infty} \exp{\left(\frac{-s^2}{4T^2}\right)} \d s \nonumber\\
    &\times& \left[ \left( s - 2i\epsilon - 2\pi ik/a\right)^{-4}
    + \frac{a^2}{6} \left( s - 2i\epsilon - 2\pi ik/a\right)^{-2} \right],
\end{eqnarray}
however, the integral in $u$ is a Gaussian integral, that is, $\int^{+\infty}_{-\infty} \exp{\left[-\left(\frac{u}{2T} + iT\Omega \right)^2\right]} \d u = \sqrt{\pi}$, and therefore we have
\begin{eqnarray}
    \mathcal{C}^\pm_{\phi^2} &=& \frac{\sqrt{\pi}}{16\pi^4} \exp{(-T^2\Omega^2)} \sum^{\infty}_{k=-\infty} \int^{+\infty}_{-\infty} \exp{\left(\frac{-s^2}{4T^2}\right)} \d s \nonumber\\
    &\times& \left[ \left( s - 2i\epsilon - 2\pi ik/a\right)^{-4}
    + \frac{a^2}{6} \left( s - 2i\epsilon - 2\pi ik/a\right)^{-2} \right].
\end{eqnarray}

It is possible to simplify even further using the following variable substitutions, given by $y \equiv s/2T$, and consequently $\d y = \d s/2T$, where these transformations have the following Jacobian $J = \d s/\d y = 2T$, aware of this, then one can write
\begin{eqnarray}
    \mathcal{C}^\pm_{\phi^2} &=& \frac{\sqrt{\pi}}{16\pi^4} \e^{-T^2\Omega^2} \sum^{\infty}_{k=-\infty} \int^{+\infty}_{-\infty} \e^{-y^2} \d y  \left[ \frac{1}{4T^2}\left( y - \frac{i\epsilon}{T} - \frac{\pi ik}{aT}\right)^{-4}
    + \frac{a^2}{6} \left( y - \frac{i\epsilon}{T} - \frac{\pi ik}{aT}\right)^{-2} \right].\nonumber\\
\end{eqnarray}
Note that it is possible to write the exponential $\exp{(-y)}$ as a Fourier transform, that is,
\begin{eqnarray}
    \e^{-y^2} = \frac{1}{\sqrt{\pi}} \int^{+\infty}_{-\infty} \e^{\xi^2} \e^{2iyk}\d\xi,
\end{eqnarray}
and with that, one obtains
\begin{eqnarray}
    \mathcal{C}^\pm_{\phi^2} &=& \frac{\e^{-T^2\Omega^2}}{16\pi^4} \int^{+\infty}_{-\infty} \d\xi\sum^{\infty}_{k=-\infty} \int^{+\infty}_{-\infty} \e^{-\xi^2} \e^{2i\xi y} \d y  \nonumber\\
    &\times& \left[ \frac{1}{4T^2}\left( y - \frac{i\epsilon}{T} - \frac{\pi ik}{aT}\right)^{-4}
    + \frac{a^2}{6} \left( y - \frac{i\epsilon}{T} - \frac{\pi ik}{aT}\right)^{-2} \right].
\end{eqnarray}

Now, the next step is to solve the integral in $y$. Note that we are faced with a complex integral, and so, to solve this integral we will use the residue theorem. According to this theorem \cite{butkov}, if $f(z)$ is analytic inside a closed contour $\gamma$ (and on this contour), except at a finite number of isolated singularities at $z = a_1, a_2, \cdots, a_n$, all located inside $\gamma$, then
\begin{eqnarray}
    \oint_{\gamma} f(z)\d z = 2\pi i \sum^{n}_{\nu=1} \mathrm{Res} f(a_\nu).
    \label{TheoremRes}
\end{eqnarray}
In this way, there is a formula that is valid for a pole of order $m$ at $z=a$ given by
\begin{eqnarray}
    \mathrm{Res} f(a_\nu) = \frac{1}{(m-1)!}\lim_{z\to a_\nu} \Bigg\{\frac{\d^{m-1}}{\d z^{m-1}}\left[ (z-a_\nu)^m f(z)\right] \Bigg\},
    \label{FormulaRes}
\end{eqnarray}
note that this expression will be useful for our integrals since they are second and fourth order pole integrals. For our case, our integrals are written as follows
\begin{eqnarray}
    \int^{+\infty}_{-\infty} \frac{\e^{2i\xi y}}{\left( y - \frac{i\epsilon}{T} - \frac{\pi ik}{aT}\right)^{2}}\d y, \\
    \int^{+\infty}_{-\infty} \frac{\e^{2i\xi y}}{\left( y - \frac{i\epsilon}{T} - \frac{\pi ik}{aT}\right)^{4}}\d y,
\end{eqnarray}
using Eq. (\ref{FormulaRes}) (for $m = 2$, $m=4$, and $a = \frac{i\epsilon}{T} + \frac{\pi ik}{aT}$) we obtain that these expressions have the respective residues, $\mathrm{Res}(m=2) = 2i\xi \exp{(-2\pi\xi k/aT)}$ and $\mathrm{Res}(m=4) = -\frac{4}{3}i\xi^3 \exp{(-2\pi\xi k/aT)}$. Thus, substituting these residues into Eq. (\ref{TheoremRes}) one obtains
\begin{eqnarray}
    \int^{+\infty}_{-\infty} \frac{\e^{2i\xi y}}{\left( y - \frac{i\epsilon}{T} - \frac{\pi ik}{aT}\right)^{2}}\d y = -4\pi\xi \exp{\left(\frac{-2\pi\xi k}{aT}\right)},  \\
    \int^{+\infty}_{-\infty} \frac{\e^{2i\xi y}}{\left( y - \frac{i\epsilon}{T} - \frac{\pi ik}{aT}\right)^{4}}\d y = \frac{8\pi\xi^3}{3} \exp{\left(\frac{-2\pi\xi k}{aT}\right)}.
\end{eqnarray}
And so our main expression becomes
\begin{eqnarray}
    \mathcal{C}^\pm_{\phi^2} &=& \frac{\e^{-T^2\Omega^2}}{24\pi^3} \int^{+\infty}_{-\infty} \xi \e^{-\xi^2}\d\xi \sum^{\infty}_{k=-\infty} \left[ \frac{\xi^2}{T^2} \exp{\left(\frac{-2\pi\xi k}{aT}\right)} + a^2 \exp{\left(\frac{-2\pi\xi k}{aT}\right)}\right].
\end{eqnarray}
The sum over $k$ is from $k = -\infty$ to $k = +\infty$, but our choice of contour for the integral subtracts the contribution to the
sum from $k \in [-\infty,0)$ leaving the result as the sum of a geometric series, and in this way we have $\sum^{\infty}_{k=0} \exp{\left(\frac{-2\pi\xi k}{aT}\right)} = \left[1 - \exp{\left(\frac{-2\pi\xi}{aT}\right)} \right]^{-1}$. Assuming these conditions, we have
\begin{eqnarray}
    \mathcal{C}^\pm_{\phi^2} &=& \frac{aT\e^{-T^2\Omega^2}}{48\pi^4} \left[ a^2 \int^{+\infty}_{-\infty} \e^{-\xi^2}\d\xi + \frac{1}{T^2}\int^{+\infty}_{-\infty} \xi^2 \e^{-\xi^2}\d\xi \right],
\end{eqnarray}
where we use the expansion $\exp{\left(\frac{-2\pi\xi }{aT}\right)} \approx 1 - \frac{2\pi\xi}{aT}$. Now, just as before, fortunately we are faced with Gaussian integrals, these integrals and their respective solutions are written as follows
\begin{eqnarray}
    \int^{+\infty}_{-\infty} \e^{-\xi^2}\d\xi &=& \sqrt{\pi}, \\
    \int^{+\infty}_{-\infty} \xi^2 \e^{-\xi^2}\d\xi &=& \frac{\sqrt{\pi}}{2}.
\end{eqnarray}
and substituting the solution of these integrals into our general expression, we obtain
\begin{eqnarray}
    \mathcal{C}^\pm_{\phi^2} &=& \frac{aT\sqrt{\pi}}{48\pi^4} \left( a^2 + \frac{1}{2T^2} \right)\e^{-T^2\Omega^2}.
\end{eqnarray}
And so, we finally obtain the final expression for the integral we wanted to solve.

\bibliography{main}
\bibliographystyle{unsrt}
	
\end{document}